\documentclass[onecolumn,aps,prb,eqsecnum,floatfix,notitlepage,10pt]{revtex4-2}

\usepackage{amsmath,amssymb,bm,euscript}
\usepackage{natbib}
\usepackage{graphicx}
\usepackage{color}
\usepackage{wasysym}
\def\LBCO-anom{La$_{1.875}$Ba$_{0.125}$CuO$_4$}

\def\LESCO{La$_{1.8-x}$Eu$_{0.2}$Sr$_x$CuO$_4$}

\def\LSCO{La$_{2-x}$Sr$_x$CuO$_4$}
\def\LBCO{La$_{2-x}$Ba$_x$CuO$_4$}
\def\YBCO{YBa$_2$Cu$_3$O$_{7-\delta}$}

\def\C60{A$_x$C$_{60}$}
\def\LNSCO{La$_{1.6-x}$Nd$_{0.4}$Sr$_x$CuO$_{4}$}

\def\SROtwo{ Sr$_{3}$Ru$_{2}$O$_{7}$}

\def\LCO{La$_2$CuO$_4$}

\def\BSCCO{Bi$_2$Sr$_2$CaCu$_2$O$_{8+\delta}$}

\def\LNSCO{La$_{1.6-x}$Nd$_{0.4}$Sr$_x$CuO$_{4}$}

\def\HgCu3{HgCa$_2$Cu$_3$O$_{8+y}$}
\def\HgCu4{HgBa$_2$Ca$_3$Cu$_4$O$_{10+y}$}
\def\TlCu{Tl$_2$Ba$_2$CuO$_{6+\delta}$}
\def\TlBCCO3{Tl$_2$Ba$_2$Ca$_2$Cu$_3$O$_{10+y}$}
\def\TlCu4{Tl$_2$Ba$_2$Ca$_3$Cu$_4$O$_{12+y}$}

\def\BiCu3{Bi$_2$Sr$_2$Ca$_{2}$Cu$_3$O$_y$}

\def\8LSCO{La$_{1.88}$Sr$_{.12}$CuO$_4$}
\def\110LNSCO{La$_{1.5}$Nd$_{0.4}$Sr$_{0.1}$CuO$_{4}$}
\def\stage4LCO{La$_{2}$CuO$_{4+\delta}$}
\def\Y248{YBa$_2$Cu$_4$O$_8$}

\def\NbSe2{NbSe$_2$}
\def\TaSe2{TaSe$_2$}
\def\TiSe2{TiSe$_2$}
\def\NaCoOH2O{Na$_{0.3}$CoO$_{2y}$H$_2$O}
\def\MgB2{MgB${}_2$}

\def\UTe2{UTe$_2$}
\def\BSNA{Ba$_{1-x}$Sr$_x$Ni$_2$As$_2$}
\def\1T-TiSe2{$1T$-TiSe$_2$}
\def\EuRbFeAs{EuRbFe$_4$As$_4$}
\def\BaFeAs122{Ba(Fe$_{1-x}$Co$_x$)$_{2}$As$_{2}$}

\begin{document}
\title{Intertwined Orders and the Physics of High Temperature Superconductors}

\author{Eduardo Fradkin}

\affiliation{ Department of Physics and Anthony J. Leggett Institute for Condensed Matter Theory, Grainger College of Engineering, University of Illinois, 1110 West Green Street, Urbana Illinois 61801, USA}

\begin{abstract}
Complex phase diagrams are generic feature of quantum materials that display high temperature superconductivity. In addition to d-wave superconductivity (or other unconventional states), these phase diagrams typically include various forms of charge-ordered phases, including charge-density-waves and/or spin-density waves, and electronic nematic states. In most cases these phases have critical temperatures comparable in magnitude to that of the superconducting state, and appear in a ``pseudo-gap'' regime. In these systems the high temperature state is not a good metal with well-defined quasiparticles but a ''strange metal''. These states typically arise from doping a strongly correlated Mott insulator. With my collaborators we have identified these behaviors as a problem with ``Intertwined Orders''. A Pair-density wave is a type of superconducting state which embodies the physics of intertwined orders. Here I  discus the phenomenology of intertwined orders and the quantum materials that are known to display these behaviors.
\end{abstract}

\maketitle

\tableofcontents
\section{Introduction}
\label{sec:intro}

I have been honored to be awarded the 2024 Eugene Feenberg Memorial Medal in Quantum Many-Body Theory ``for pioneering applications of quantum field theory to the understanding of emergent, many-body physics of quantum systems, in particular composite fermions, and electronic liquid crystalline and pair density wave phases of correlated electronic systems.'' 

This paper is based on the talk that I gave at the 2024 Conference on Recent Progress in Many-Body Theories held at the University of Tsukuba (Tsukuba, Japan) on September 23-27, 2024 on the topic of Intertwined Orders and the Physics of High Temperature Superconductors. Here I summarize the contents of my talk in which I presented my view on the current status of our understanding of the problem of high-temperature superconductivity in strongly correlated electronic systems. This perspective is the result of work that I have done for over two decades on this central problem in Condensed Matter Physics. Although I have worked in several other areas of Condensed Matter Physics such as the fractional quantum Hall effects and in the gauge theory approaches to strongly correlated systems, areas in which the field theory approaches play a fundamental role, this paper (as was my Tsukuba talk) is focused on my contributions to the problem of Intertwined Orders in High Temperature Superconductors.  I have presented a comprehensive presentation of my work and perspective of these other subjects in  my book \textit{Field Theories of condensed Matter Physics, Second Edition} \cite{Fradkin-FTCMP-2013}.

\section{What is a Superconductor}
\label{sec:what}

The resistivity of most metals vanishes below some sufficiently low temperature $T_c$. This phenomenon is known as \textit{superconductivity} and hence can conduct electricity without dissipation. It was discovered by Heike Kamerlingh Onnes in 1911 when he cooled down mercury to below 4.2 degrees Kelvin. In subsequent years many superconductors were discovered. Until 1986 the critical temperature of known superconductors was typically below 20 degrees Kelvin. 

\begin{figure}[hbt]
\center
\includegraphics[width=0.45\textwidth]{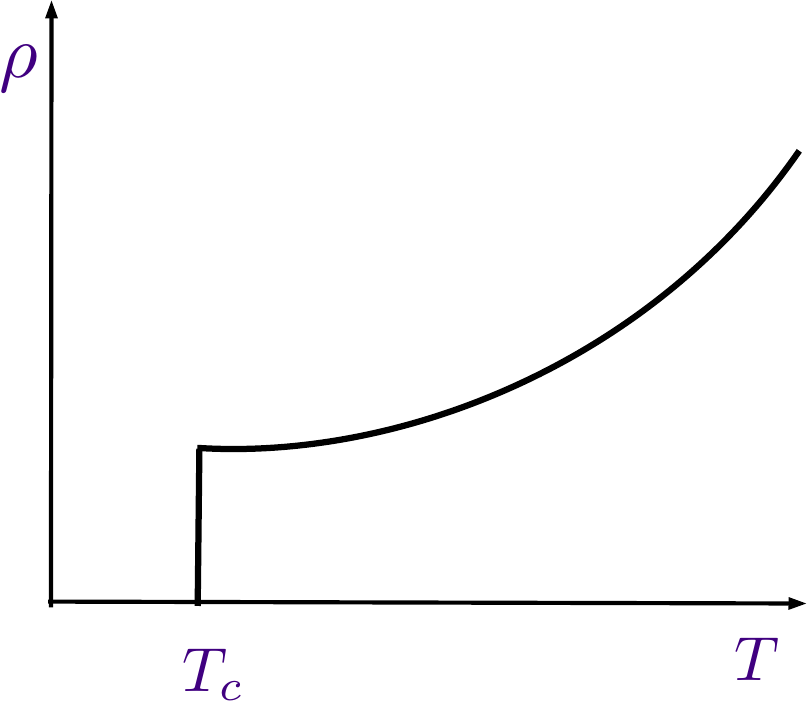}
\caption{Schematic temperature-dependence $\rho(T) \propto{T^2}$  of the resistivity of a simple metal above the superconducting transition.}
\label{fig:normal-resistivity}
\end{figure}

The high temperature phase of such ``conventional'' superconductors is a normal metal.  In a normal metallic state the resistivity has a quadratic temperature dependence, $\rho(T) \propto T^2$, as shown in Fig.\ref{fig:normal-resistivity}.In a simple metal  the mean-free-path of the conduction electrons at low enough temperatures is large compared to their mean separation. In a metal static electric fields are screened and vanish deep in the bulk of the metal. Normal metals of this type are well described by the Landau Theory of the Fermi Liquid \cite{Landau1957,pines-nozieres-1966,Baym1991}.

Below a critical temperature $T_c$ the superconducting state appears and the longitudinal resistivity vanishes and currents flow without dissipation. In the superconducting state external static magnetic fields, below some critical field $H_c$, are expelled from the bulk of the material and penetrate up to a scale $\lambda$ known as the penetration depth. This is the Meissner effect. It implies that in the superconducting state photons are gapped (and behave as ``massive particles'') and electromagnetic fields cannot propagate. In contrast, in a metal electromagnetic fields are overdamped. In conventional superconductors fermionic quasiparticles are fully gapped. However in ``unconventional'' (anisotropic) superconductors have ``nodal'' fermionic excitations whose gap vanishes at special places on the Fermi surface.

\subsection{Landau Fermi Liquids}
\label{sec:Landau}

The Landau Theory of the Fermi Liquid \cite{Landau1957} describes a system of interacting fermions (electrons in the case of interest here) at finite density $n$ and chemical potential $\mu$. In the non-interacting limit all one-particle states with energy $\varepsilon(\bm p)$ less that the Fermi energy  $E_F\equiv\mu$  are occupied and all states with energy higher that the Fermi energy are unoccupied, and the occupation number of single particle states has a discontinuity (a jump) at the Fermi energy. 

The fundamental assumption of the Landau Theory  is that interactions \textit{renormalize} the electronic states and become \textit{quasiparticles} with the \textit{same} quantum numbers (charge, spin, etc) as ``bare'' electrons. The electronic quasiparticles have an effective mass $m^*$. In a dense Fermi Liquid the Coulomb interactions are screened, the quasiparticles have weak and short-range forward scattering interactions parametrized by Landau parameters, and the discontinuity in the single-particle occupation number at the Fermi energy is renormalized down to a value $Z<1$ (known as the quasiparticle residue). At low energies (i.e. close to the Fermi energy) the electron propagator $G_{\sigma \sigma'}(\omega, \bm p)$ at low energies has a quasiparticle pole of the form
\begin{equation}
G_{\sigma \sigma'}(\omega, \bm p)= \frac{Z \,}{\omega-\varepsilon(\bm p)+i\Sigma^{''}(\omega, \bm p)} \delta_{\sigma, \sigma'}+\ldots
\label{eq:electron-propagator}
\end{equation}
where $\varepsilon(\bm p)=v_F (|\bm p|-k_F)$ and $\Sigma^{''}(\omega) \propto \omega^2$, the imaginary part of the self-energy $\Sigma(\omega, \bm p)$,  is the \textit{quasiparticle width}, the decay rate $\Gamma(\omega)$. In the Landau theory the quasiparticle width is small compared to the energy, 
$\Gamma(\omega) \propto \omega^2 \ll |\omega|$, and the quasiparticles states are well defined at low energy. The ellipsis in Eq.\eqref{eq:electron-propagator} denotes the contribution of the multi-particle continuum.

The pole structure of the propagator implies that in the Fermi liquid there is a Fermi surface at $|\bm p|=k_F$ (here for simplicity we assumed an isotropic system) where the quasiparticles are long-lived. The analytic structure of the propagator also implies, among other things,  that the resistivity has a $T^2$ temperature dependence \footnote{Naturally, the resistivity is finite only in the presence of momentum (and/or energy) relaxation mechanisms such as umklapp processes, disorder, phonons, etc. The temperature dependence of the resistivity is different for different processes. The $T^2$ dependence is for  a Fermi liquid with some weak relaxation mechanism, see Ref. \cite{Baym1991}.}. The existence of the Fermi surface is documented  in different types of experimental probes such as quantum oscillations   of resistivity, magnetization and electronic specific heat in magnetic fields, and in angle-resolved photoemission (ARPES) experiments.

\begin{figure}[hbt]
\center
\includegraphics[width=0.45\textwidth]{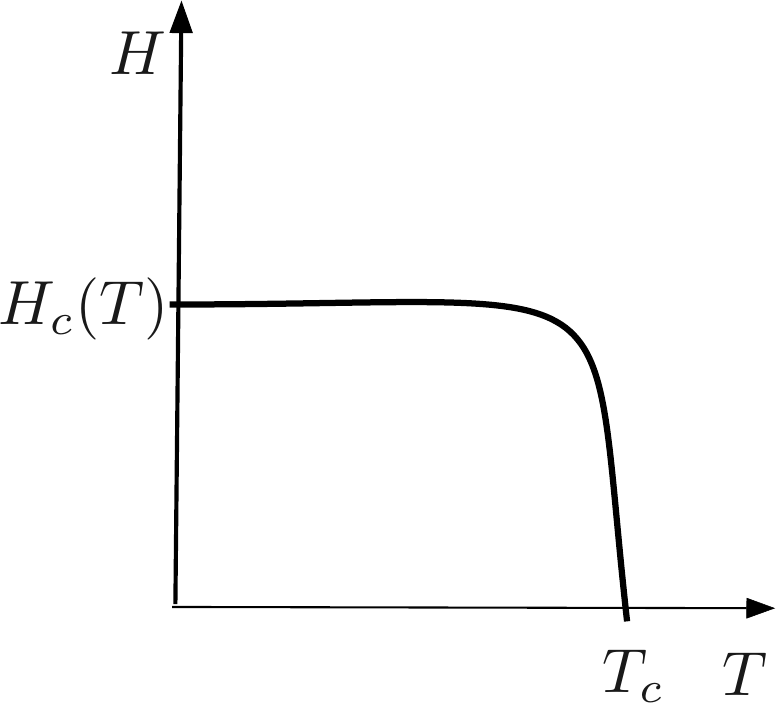}
\caption{The Meissner effect: qualitative temperature-dependence of the critical magnetic field $H_c(T)$ of a superconductor.}\label{fig:critical-field}
\end{figure}   

\subsection{Ginzburg-Landau Description of the Superconducting State}
\label{sec:theory}

At a phenomenological level, the thermodynamic phase transition to a superconducting state is described by a Ginzburg-Landau (GL) theory \cite{deGennes}. In this theory the superconductor is characterized by a \textit{complex order parameter field}, which we will denote by $\Delta(\bm x)$.  Close to $T_c$ the free energy density of the superconductor has the form
\begin{equation}
\mathcal{F}=\kappa \left|\left(\bm \nabla +i \, \frac{2e}{\hbar c} {\bm A}(\bm x)\right)\Delta(\bm x) \right|^2+r(T)|\Delta(\bm x)|^2+\frac{u}{2} |\Delta(\bm x)|^4
\label{eq:GL}
\end{equation}
where $\kappa \propto \rho_s$ is the stiffness of the order parameter (here $\rho_s$ is the superfluid density), and determines the magnitude of the penetration depth $\lambda$; $r(T)\equiv\xi(T)^{-2}$ is the inverse correlation length (squared) and $u>0$ is an interaction which makes the system thermodynamically stable.
Close to $T_c$ the quadratic term changes sign $r(T) \propto (T-T_c)$ signaling a continuous phase transition into the superconducting state. 

In the absence of an external magnetic field, the Ginzburg-Landau free energy is invariant under the \textit{global} $U(1)$ symmetry transformation
\begin{equation}
\Delta(\bm x) \mapsto \exp(2i \theta)\, \Delta(\bm x)
\label{eq:globalU1}
\end{equation}
where $\theta$ is a constant phase.
In the presence of an external magnetic field, denoted by the vector potential $\bm A(\bm r)$, the $U(1)$ becomes a \textit{local}  symmetry (or \textit{gauge} symmetry)
\begin{equation}
\Delta(\bm x) \mapsto \exp(2i \theta(\bm x))\, \Delta(\bm x),\quad \bm A(\bm x) \mapsto \bm A(\bm x)-\frac{\hbar c}{e} \bm \nabla \theta(\bm x)
\label{eq:gaugeU1}
\end{equation}
The factor of $2$ in Eqs.\eqref{eq:GL}, \eqref{eq:globalU1} and \eqref{eq:gaugeU1} anticipates the fact the superconducting order parameter is a condensate of Cooper pairs and, hence, has to transform as a charge $2$ complex field.

The low temperature behavior of superconductors in magnetic fields depends on the value of the ratio $\kappa\equiv \lambda/\xi$: for  $\kappa<1$ (``type I'') the magnetic field is fully expelled below $H_c$ while for $\kappa>1$ (``type II'') the full Meissner state exists below a critical field $H_{c1}$, for intermediate larger fields up to  an upper critical field $H_{c2}$ there is a mixed state in which the external field penetrates the bulk of the superconductor in the form of a lattice of quantized (Abrikosov) vortices, each carrying a magnetic  flux $hc/2e$.

\subsection{BCS Theory}
\label{sec:bcs}

For many years  the physical origin of superconductivity defied a microscopic explanation. This changed in 1957 when John Bardeen, Leon Cooper and J. Robert Schrieffer (BCS) developed the first successful microscopic theory of superconductivity \cite{BCS,Schrieffer1964}. 

The central assumptions of BCS theory are  1) that the high temperature (``normal'') state is a good metal, well described by the Landau Theory of the Fermi Liquid and 2) that close to the Fermi surface the electronic quasiparticles experience an \textit{effective attractive interaction}. In the BCS picture the weak attractive interaction leads to the formation of a weak bound state of two electrons known as \textit{Cooper Pairs}  which behave as a boson and thus the ground state is a \textit{condensate} of Cooper pairs with total zero momentum. In BCS theory the effective attractive interaction is assumed to be due to the exchange phonons. In other words, in BCS the superconducting state is a  of Cooper pairs each carrying a $2e$ electric charge. 

Being a charge condensate the superconducting state has profound implications for its electromagnetic response such as the expulsion of a static magnetic field, the Meissner effect. This also means inside a superconductor photons behave as massive particles. For these and other reasons, BCS theory (and the Ginzburg-Landau version) has had tremendous impact in other areas of physics and, particularly, in high energy and nuclear physics, such as the Nambu-Jona-Lasinio model of dynamical symmetry breaking \cite{Nambu-1961}, the work by Peter Higgs (and others) on what is now called the Higgs mechanism \cite{Higgs-1964}, the Weinberg-Salam theory of weak interactions \cite{Weinberg-1967,Salam-1968}, the work by Susskind (and others) to formulate a theory of a ``composite Higgs'' field \cite{Susskind-1979}, and the theory of color superconductivity in dense hadronic matter \cite{Alford-2001a,Alford-2001b}. 

While in BCS theory of superconductivity the Cooper pairs are assumed to be in spin singlet state, in liquid $^3$He the superfluid $A$ and $B$ phases are condensates of Cooper pairs in a spin triplet state,  and the effective attractive interaction is  the exchange of paramagnons \cite{Leggett1975}.
\begin{figure}[hbt]
\center
\includegraphics[width=0.40\textwidth]{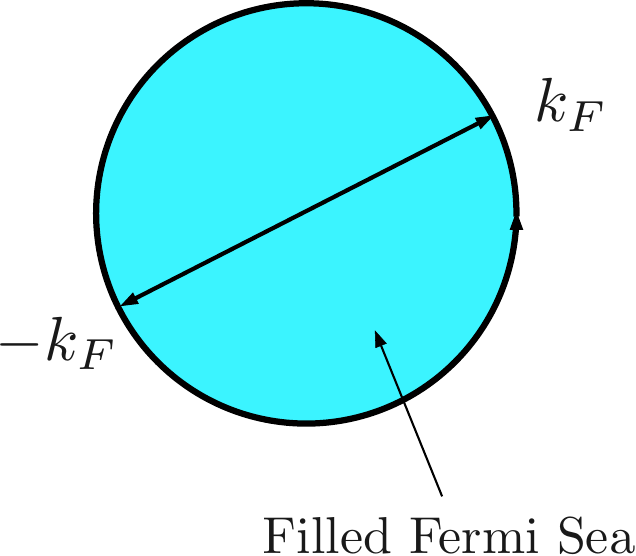}
\caption{Cooper pair.}
\label{fig:cooper}
\end{figure}   

More formally, the order parameter of a superconducting state, which we denoted by $\Delta(\bm k)$, is the expectation value of the \textit{pair field operator} 
\begin{equation}
\hat \Delta(\bm k)=c^\dagger_\downarrow(\bm k) c^\dagger_\uparrow(-\bm k)
\label{eq:pair-field}
\end{equation}
where $c_{\sigma}^\dagger (\bm k)$ creates an electron with spin $\sigma=\uparrow,\downarrow$ and momentum $\bm k$, which is a wave vector close to the Fermi surface of the metallic state. In an isotropic system $|\bm k|\cong k_F$. For simplicity, here we assumed that the pair field is a spin singlet. 

Thus, if the ground state expectation value of the pair field operator is non-vanishing then this state represents a superconductor. Alternatively, the ground state is a superconductor if the equal-time correlator of the pair field in position space satisfies 
\begin{equation}
\lim_{\bm R \to \infty}\Big< \hat \Delta(\bm r +\bm R)^\dagger \Delta(\bm r - \bm R) \Big>\equiv |\Delta|^2 \neq 0
\label{eq:ODLRO}
\end{equation}
where $\hat \Delta(\bm r)$ is the Fourier transform of the pair field operator of Eq.\eqref{eq:pair-field}.
This condition, known as \textit{off-diagonal long range order}, implies that in the superconducting state the global $U(1)$ symmetry is spontaneously broken and that electric charge is conserved only modulo two.

The effective Hamiltonian for a BCS state is the pairing Hamiltonian (for spin singlet superconducting state) \cite{Schrieffer1964}
\begin{equation}
H_{\rm BCS}=\int_{\bm k}\Big\{ \sum_\sigma (\varepsilon(\bm k)-\mu) c_\sigma^\dagger (\bm k) c_\sigma (\bm k)+ \Delta(\bm k) c^\dagger_\downarrow(\bm k) c^\dagger_\uparrow(-\bm k)+h.c. +\frac{|\Delta(\bm k)|^2}{g}\Big\}
\label{eq:BCS-H}
\end{equation}
where $\bm k$ is restricted to a small shell of momenta close to the Fermi surface and $g$ is a dimensionless coupling constant for a short-range attractive interaction. The spectrum of states of the Hamiltonian of Eq.\eqref{eq:BCS-H} are not electrons but  linear combinations of electrons and holes called \textit{Bogoliubov quasiparticles}. At low energies Bogoliubov quasiparticles are fermions but do not carry charge which is carried by the condensate. In this sense all fermionic quasiparticles of superconductors are Majorana fermions since they obey a charge neutrality condition.

When  $\Delta(\bm k)\neq 0$ the ground state of the Hamiltonian of Eq.\eqref{eq:BCS-H}  is a superconducting state. At zero temperature the quasiparticle spectrum of this state has an energy gap $2\Delta(\bm k)$. For an isotropic superconducting state $\Delta(\bm k)=\Delta$ is isotropic and has the form
\begin{equation}
\Delta= \omega_0 \exp\left(-\frac{\textrm{const}.}{g}\right)
\label{eq:gap}
\end{equation}
where $\omega_0$ is an energy scale which in BCS theory is the Debye frequency of the phonons. 

The expectation value of the pair field operator $\Delta(\bm k)=\langle \hat \Delta(\bm k) \rangle$ is the order parameter of the superconducting state. In position space $\Delta(\bm r)$ transforms as a charge 2 complex scalar field as it is assumed in Ginzburg-Landau theory. In addition the order parameter may also transform non-trivially under spatial rotations. In the simplest case $\Delta(\bm k)$ is isotropic and the symmetry is $s$-wave.
In anisotropic superconductors, such as the copper-oxide high temperature superconductors, which are strongly layered materials, the gap is anisotropic and has a $d$-wave symmetry
\begin{equation}
\Delta_{\rm d wave}(\bm k) \sim \Delta_0 \, (k_x^2-k_y^2)
\label{eq:dwave}
\end{equation}
In this case the spectrum is gapless at the values of $\bm k$ at four points in momentum space where the lines $k_x=\pm k_y$ intersect the Fermi surface. Near these ``nodal'' points the spectrum has a massless  Dirac form. The quantity $\Delta_0$ is gap scale away from the nodal point.  In the case of the superfluid phases of $^3$He the order parameter has $p$-wave symmetry \cite{Leggett1975}. Superconducting order parameters with more exotic symmetries, such as topological superconductors, have also been considered  such as chiral superconductivity with symmetry $p_x\pm ip_y$.

A crucial feature of Eq.\eqref{eq:gap}  is that $\Delta\neq 0$ no matter how small is the coupling constant $g$ and that the non-analytic dependence on $g$ is an essential singularity. This non-perturbative behavior implies that the metal has an instability to a superconducting state for arbitrarily weak attractive interactions. On the other hand, in BCS theory the critical temperature is proportional to the superconducting gap at zero temperature, $T_c \propto \Delta$. As a result the critical temperature of BCS superconductors is necessarily a small fraction of the scale $\omega_0$. Hence BCS theory necessarily applies only to low temperature superconductors. For these reasons typical values of $T_c$ for conventional superconductors is low. the most studied examples, aside from mercury, are niobium (with a $T_c \sim 10^\circ$K) and Nb$_3$Ge with a $T_c\sim 23^\circ$K). Niobium is the most widely used superconductor in applications.

\section{High Temperature Superconductors are Different}
\label{sec:htsc}

Using diverse arguments, prior to 1986 it was widely believed that the critical temperature of superconductors could not be higher than about $25^\circ$K. In 1986 high temperature superconductivity was discovered \cite{Bednorz-1986} in several copper oxide materials, first in {\LBCO} and  {\LSCO} (with a $T_c \sim 38^\circ$K) and {\YBCO} (with a $T_c=92^\circ$K). The critical temperature of other copper-oxides, such as in the Thalium compounds, is even higher,  $T_c\sim 150^\circ$K (at ambient pressure). Other superconductors with comparable (but not as high) have been discovered, notably the iron-based superconductors. Recent studies in hydride materials with comparable (and higher) values of $T_c$ (at extremely high pressure) have been reported (not as well studied and not well understood).

The case of the copper-oxide superconductors was, and remains, particularly startling. In all cases the parent compound is a Mott insulator and superconductivity is achieved by doping. In all cases these materials are strongly layered (some more than other) and they are all made of layers of CuO$_2$ separated by other elements. For instance, the parent compound of the lanthanum series of copper oxide superconductors is {\LCO}. In the parent material {\LCO} the band close to the Fermi energy has one electron at each copper atom in the unit cell. Band theory predicts that it should be a metal. Density functional theory calculations also predicted metallic behavior. However, {\LCO} is a Mott insulator meaning that it has a large ``on-site'' Coulomb interaction (the ``Hubbard $U$'') leading to an insulating state with a large charge gap of the order of 1eV. Moreover, the Mott insulator is actually an antiferromagnet with a well developed N\'eel state, which develops at about $350^\circ$ K,  and a magnetic moment of about 1/2 of the Bohr magneton per copper site. 

\begin{figure}[hbt]
\center
\includegraphics[width=0.50\textwidth]{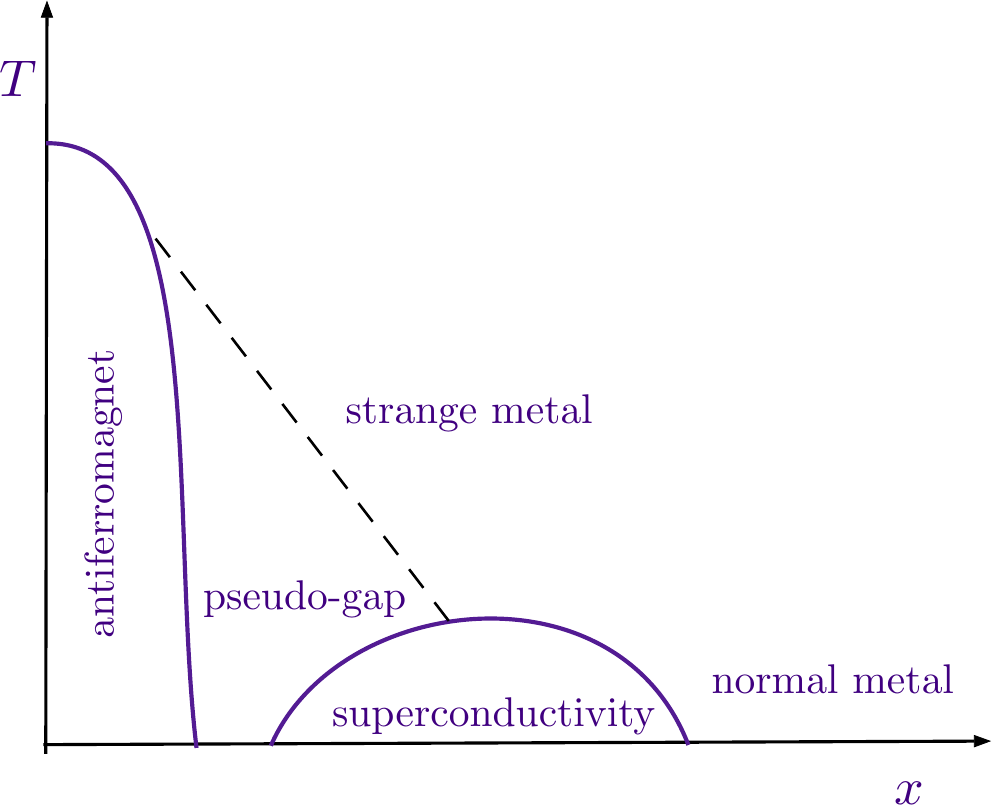}
\caption{Schematic phase diagram for a copper-oxide superconductor. The regions are not drawn at scale. Here $T$ is the temperature and $x$ is the (hole) doping.}
\label{fig:schematic-htsc}
\end{figure}

Doping {\LCO} with barium and/or strontium results in a relatively small number of mobile holes. The mobile holes drive the system into a superconducting state whose critical temperature $T_c$ has the shape of a dome as a function of doping with its maximum at about $15\%$ doping. Above the maximum superconducting $T_c$ is the strange metal phase and in-between there is the ``pseudo-gap'' regime. for large dopings, to the right of the superconducting dome there is a metallic phase widely believed to be a normal Fermi liquids. This pattern is present in the phase diagrams of all the copper-oxide superconductors, shown schematically in Fig.\ref{fig:schematic-htsc}. 

In addition to proximity to a Mott insulating parent compound, which is a Mott insulator with long range N\'eel antiferromagnetic order, the phase diagram has many other distinctive features. One is the ``strange metal'' phase above the maximum $T_c$ (the top of the superconducting dome). This metal is strange in the sense that in this phase the resistivity has a linear temperature dependence over many decades, unlike normal metals where it is quadratic, c.f. Fig.\ref{fig:normal-resistivity}. The linear temperature dependence of the resistivity is a violation of the Landau theory of the Fermi liquid. This behavior is also unconventional in that as the temperature increases there is a point at which the nominal mean free path becomes shorter than the mean spacing between the charge carriers which is why they are also called ``bad metals'' \cite{Emery-1995}. This limit is the so-called Ioffe-Regel criterion and  is violated in all the copper-oxide superconductors (and in many other materials).

 These effects imply that the picture of the low energy electronic states as being described weakly coupled quasiparticles fails. One way it can fail is if the quasiparticle width (the decay rate) is large compared to the energy. Raman scattering \cite{Cooper-1988}, angle-resolved photoemission \cite{Damascelli-2003},  and momentum-resolved electron energy-loss experiments \cite{Vig-2017,Abbamonte-2025} have given strong hints that the quasiparticle picture may not apply in this regime and that these systems are in a strongly interacting regime. 
 
A feature of conventional metals is the observation of quantum oscillations. In a non-interacting system of fermions in the presence of  an external uniform magnetic field quantum oscillations occur as the Landau levels cross the Fermi energy. This effect is broadly used as a way to map out the Fermi surface. In weakly interacting systems the quantum oscillations in the orbital magnetization are well described by the Lifshitz-Konsevich formula. In the high temperature superconductors such as {\YBCO} quantum oscillations are seen only if the magnetic field is stronger than a value $H~18$ Tesla. It is often stated that the observed quantum oscillations reveal the underlying Fermi surface hidden by the superconducting state. However this is almost surely incorrect since at this value of the magnetic field a thermodynamic phase transition is observed in the form of an anomaly in the ultrasound velocity. In other words, the magnetic field is not revealing an underlying Fermi liquid but is stabilizing a phase of the system which competes with the superconductivity. Below we will argue that this phase has a charge-density-wave (or ''charge stripes'').

 To this date there isn't  good theory of the strange metal phase and its explanation remains one of the challenges to Condensed Matter Theory. A commonly used theoretical proposal is to assume that there is a quantum critical point somewhere under the superconducting dome. Although this is an appealing proposal it has problems. One is that it is hard to find a quantum critical theory which naturally explains the linear resistivity. The other problem is that this behavior is seen not just in close proximity to a quantum critical point but in bread ranges of parameters and in different materials. It is hard to imagine that there will always be a suitable quantum critical point available in all cases. The linear temperature dependence of the resistivity can also be explained by a scattering rate that is proportional to the temperature. An appealing hypothesis, known as ``Planckian dissipation'' \cite{Hartnoll-2022}, states that the scattering rate is universal and given just $k_BT/\hbar$ where $k_B$ is the Boltzmann constant and $\hbar$ is the Planck constant. This picture would suggest that in the strange metal enjoys some sort of scale invariance up to arbitrarily high temperatures. While this scenario is natural  in the AdS/CFT picture, it is unclear why it should apply. Recent work on generalizations of the SYK model have  clarified the connection between the AdS/CFT theories and strange metals (for details see Ref. \cite{Esterlis-2025}). 

On the other hand, the superconducting state has $d$-wave symmetry \cite{Wollman-1993,Kirtley-1995} which is consistent with a system having strong local repulsive interactions instead of weak attractive interactions. The high temperature materials are strong type II superconductors with values of $\kappa \sim 100$. Thus, the zero-temperature superconducting coherence length is two orders of magnitude smaller than the penetration depth. All of these behaviors are indications of strong interaction physics. Since the penetration depth is of the order of one micron, the coherence length is a few lattice spacings. However, aside from that, the superconducting state itself looks quite ``normal'' in the sense that its  Bogoliubov quasiparticles are sharply defined. Clearly, contrary to the BCS picture, in the high temperature superconductors the Bogoliubov quasiparticles are not what is left of the quasiparticles of a preexisting metallic state after pairing sets in but, instead, are quasiparticles that are only well defined in the superconducting state. The existence of Bogoliubov quasiparticles with a Dirac spectrum has been evinced in angle-resolved photoemission \cite{Damascelli-2003} and scanning tunneling microscopy and spectroscopy experiments in {\BSCCO} \cite{Hoffman-2002}, and in thermal conductivity experiments in {\YBCO} \cite{Aubin-1997}. 

Nevertheless,  in spite of the strongly interacting nature of these systems, the superconducting state of these systems is still described by a pair-field order parameter since they have the same spontaneously broken symmetry as in any other superconducting state. What is different here is the way the pair-field transforms under the point group symmetries of the crystal.

The existence of the pseudo-gap regime is another startling difference with the physics of conventional superconductors. A key feature of the pseudo-gap is the existence in addition to superconductivity of other ordered phases such as spin-density waves (SDW) (often called ``spin stripes'') \cite{Tranquada-1995}, charge-density waves (CDW) (``charge stripes'') \cite{Abbamonte-2005,Hayden-2024}, both of which exhibit periodicities of just a few lattice constants, and nematic order \cite{Hinkov-2008,Chang-2011}. The existence of these phases, often with critical temperatures of comparable magnitude, led to the phase diagrams of these systems to be rather richer (and more complex) than in conventional superconductors. 

In some materials these orders are static, such as in the Lanthanum series of the copper oxides, or  are seen as ``fluctuating orders'' (as seen in {\YBCO}), meaning that they are observed on low energy fluctuations and are proximate to an ordered state \cite{Kivelson-2003}. However, all materials have disorder of varying types and degrees, which couple strongly to phases of matter that break translation and/or rotation invariance (or point group symmetries) in the form of effective random fields that couple linearly to the associated order parameters. However, it has long been established, from the pioneering work of Y. Imry and S.-K. Ma \cite{Imry-1975} and of K. Efetov and A. Larkin \cite{Efetov-1977}, that such types of disorder destroys long range order below four spatial dimensions for continuous symmetries or two dimensions for discrete symmetries. Thus, to varying degrees, systems with a tendency to have these type of phases never truly exhibit long range order.

As we will discuss in next sections there are two possible ways to regard this complexity. One alternative is to think that these orders compete with each other and the other alternative is to regard all of the orders as being part of the same physics and hence that they are ``intertwined'' with each other. For reasons that I explain below I advocate  the second viewpoint.

\section{Electronic Liquid-Crystal Phases}
\label{sec:elc}

The discovery of the high temperature superconductors, and the realization that they cannot be understood in the context of a weak coupling theory such as BCS, motivated a reexamination of its possible mechanisms. These investigations, in turn, led to the realization of the possible existence many novel phases of matter which had not been expected before 1986. 

As we said before,  high temperature superconductors arise from doping a  material which is a Mott insulator. In the copper-oxide superconductors the Mott insulator is a N\'eel antiferromagnet with a critical temperature $T_N \sim 300$K with ordering wave vector $\bm Q=(\pi, \pi)$. In 1995 John Tranquada and coworkers discovered  that the copper oxide {\LNSCO} at  doping near $x=1/8$ exhibits static incommensurate spin order with unidirectional order with wave vector $\bm Q(\delta)=(\pi-\delta,\pi-\delta)$ with an associated periodic modulation of the local charge distribution with wave vector $\bm K=2\bm Q(\delta)$ \cite{Tranquada-1995,Tranquada-1996}. In {\LNSCO} the charge order has period $4$ and was consistent with charges accumulating at anti-phase domain walls of the spin order. It took more than ten years to understand the wider significance of this discovery.

In 1998 Steven Kivelson, the late Victor Emery and I looked  at the implications of the discovery of spin and charge stripes. The result of our work was the formulation of the existence of novel phases of electronic matter which we called \textit{electronic liquid crystal phases} \cite{Kivelson-1998}, that qualitatively resembled the phases of \textit{classical} liquid crystals \cite{Chaikin-1995,deGennes-LC}.

In 1993 Emery and Kivelson \cite{Emery-1993} showed that doping a  Mott  insulator led to a tendency of the doped holes to phase-separate into hole-rich regions while the repulsive Coulomb interactions frustrated this process. As a result, at low doping, where screening is essentially absent, the doped holes tend to form a \textit{crystal}. In a continuum system these crystal is essentially a Wigner crystal and has the structure of a triangular lattice. However, in the copper-oxide materials the square lattice plays a strong role and the crystal is naturally a square lattice. 

\begin{figure}[hbt]
\center
\includegraphics[width=0.55\textwidth]{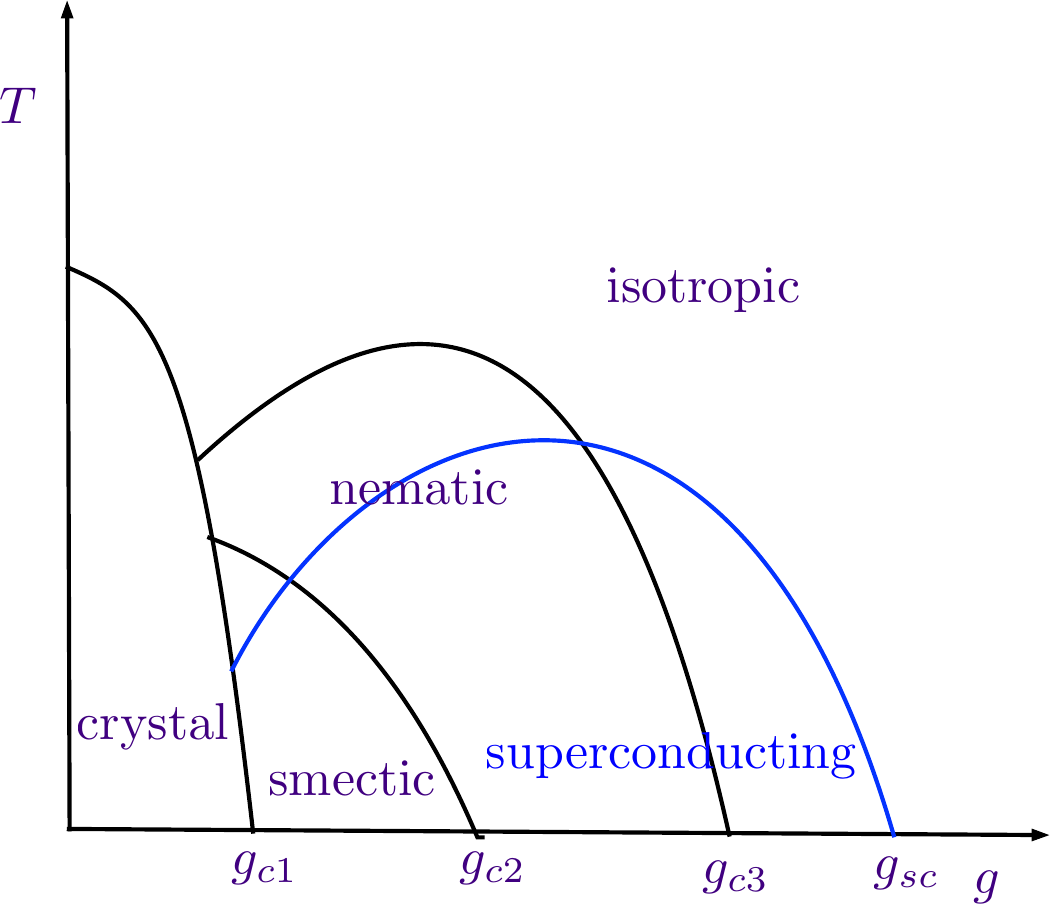}
\caption{Schematic phase diagram of electronic liquid crystal phases in a doped Mott insulator. The regions are not drawn at scale. Here $T$ is the temperature and $g$ is a ``coupling constant'' (qualitatively representing doping) which measures the strength of quantum fluctuations. See text for details.}
\label{fig:elc-phase-diagram}
\end{figure}

Building on the Emery-Kivelson result, in 1998 with Kivelson and Emery \cite{Kivelson-1998} we showed that the crystal phase can melt either by effects of thermal fluctuations or by quantum fluctuations. Since these systems are (quasi) two-dimensional we argued, by analogy with the classical theory of melting in two dimensions \cite{Halperin-1978,Young-1979}, that the crystal phase can melt first into a \textit{smectic phase}, followed by a transition to a \textit{nematic phase} and finally to an \textit{isotropic phase} (see Fig.\ref{fig:elc-phase-diagram}). In the present context the \textit{crystal phase} is an insulating state which breaks spontaneously the translating invariance of the square lattice in both directions (i.e.a square crystal), the \textit{smectic} is an unidirectional state of charge order (a stripe phase which breaks translation invariance in only one direction), while the \textit{nematic} is an anisotropic  translationally-invariant phase which breaks spontaneously the discrete $C_4$ point group symmetry of the square lattice. Each one of these thermal phase transitions are of the Kosterlitz-Thouless universality class \cite{Kosterlitz-1973}. On the other hand, these phases may also melt by a sequence of \textit{quantum} phase transitions, denoted by $g_{c1}$, $g_{c2}$ and $g_{c3}$ 
in Fig.\ref{fig:elc-phase-diagram}.

\begin{figure}[hbt]
\center
\includegraphics[width=0.4\textwidth]{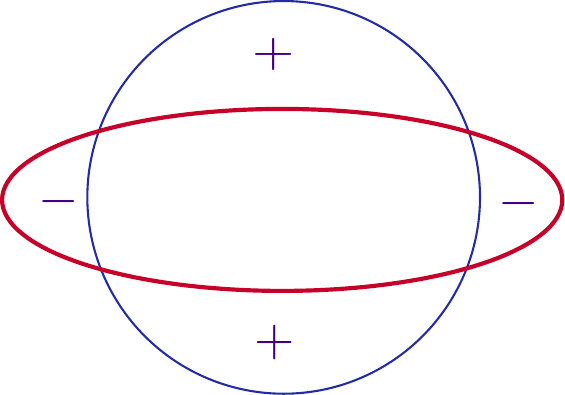}
\caption{Spontaneous distortion of the Fermi surface and quadrupolar charge distribution at a Pomeranchuk instability of a Fermi liquid into en electron nematic phase; the Fermi surface of the isotropic fluid is shown in blue and the anisotropic Fermi surface of the electronic nematic fermionic fluid is shown in red. Here $-$ shows states with excess of electrons and $+$ states with excess of holes. }
\label{fig:nematic}
\end{figure}

While crystalline and stripe (or CDW) phases can be obtained theoretically using conventional Hartree-Fock type approaches, nematic phases are more intrinsically strong coupling states. Several physical mechanisms can lead to a nematic ground states. Sun and coworkers \cite{Sun-2008} showed  that an electronic nematic state can be reached  by quantum melting a stripe phase  for example as described a quantum version of the McMillan-deGennes theory of the nematic-smectic transition in liquid crystals \cite{Chaikin-1995}. Oganesyan, Kivelson and  I showed that  at the  Pomeranchuk instability in the quadrupolar channel ($\ell=2$) of two-dimensional Fermi liquid there is a quantum phase transition to a nematic state \cite{Oganesyan-2001}. In the nematic phase gets spontaneously distorted breaking rotational invariance mod $\pi$, see Fig.\ref{fig:nematic}.  In continuum systems the nematic \textit{phase} is a non-Fermi liquid but lattice effects render it into a conventional Fermi liquid, expect at the quantum phase transition. An electron nematic was also shown to arise in the low doping limit of the Emery (three-band) model of the copper oxide lattice in the strong coupling regime \cite{Kivelson-2004}.

\begin{figure}[hbt]
\center
\includegraphics[width=0.35\textwidth]{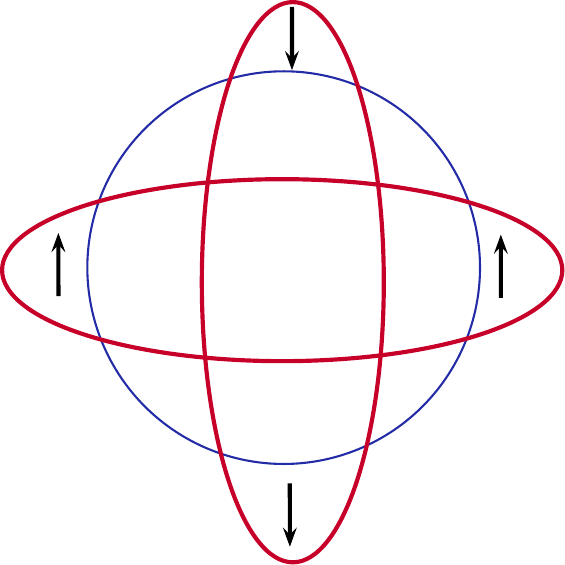}
\caption{Quadrupolar spin polarization of the nematic-spin-nematic state, now known as  an ``altermagnet''. Here $\uparrow$ and $\downarrow$ denote states with different spin polarization.}
\label{fig:nematic-spin-nematic}
\end{figure}

On the other hand, Congjun Wu, Kai Sun, Shoucheng Zhang and I showed that a Pomeranchuk instability in the spin triplet channel leads to generalizations of the nematic phase in the spin channel \cite{Wu-2007}. One example is the ``nematic-spin-nematic'' phase, conjectured in Ref.\cite{Kivelson-2003} and shown in Fig.\ref{fig:nematic-spin-nematic}. This state has d-wave symmetry: is invariant under a $\pi/2$ spatial rotation followed by a spin flip. This state has the same symmetry of the \textit{altermagnet} states which are being widely discussed in the current literature \cite{Smejkal-2022}. Much as its charge cousin, these spin triplet particle-hole condensates at zero total momentum cab also have higher angular momentum. In Ref.\cite{Wu-2007} it is also shown that there is a second class (the ``beta-phase'' in the terminology of Ref. \cite{Wu-2007}) of anisotropic phases in the triplet channel in which the spin polarization \textit{winds} twice (for $\ell=2$) around the Fermi surface with a non-vanishing Berry phase. The Pomeranchuk instability is one of the possible \textit{mechanisms} for the formation of these states (just as it is for the nematic) but it is natural in the continuum description of the Landau Fermi liquid. However, in most materials that display altermagnetism the lattice (and its symmetries) plays a central role and other mechanisms are possible as well. Nevertheless all altermagnets have the same symmetry breaking as the d-wave spin triplet state of Ref.\cite{Wu-2007} and, in this sense, they are the same state with the caveat that many altermagnetic materials are insulating whereas the Pomeranchuk counterpart is a metal. A discussion of the similarities and differences can be found in Ref.\cite{Jungwirth-2025}.

Electronic liquid crystal phases can be diagnosed by order parameters similar to those of their classical counterparts. Let us consider an incommensurate charge-ordered state on a square lattice which has $C_4$ point group symmetry. In this case the charge distribution can be modulated along one or both symmetry directions with orthogonal ordering wave vector $\bm Q_1$ and $\bm Q_2$ (with $\bm Q_1 \cdot \bm Q_2=0$). The charge density profile is
\begin{equation}
\rho(\bm r)=\rho_0+\sum_{i=1,2} \Big[\rho_{\bm Q_i}(\bm r) \exp(i \bm Q_i \cdot \bm r)+\rho_{-\bm Q_i}(\bm r) \exp(-i \bm Q_i \cdot \bm r)\Big]
\label{eq:CDW}
\end{equation}
Since the local charge density $\rho(\bm r)$ is a hermitian operator we must have  $\rho_{\bm Q_i}(\bm r)=\rho_{-\bm Q_i}^\dagger(\bm r)$. If the expectation values of  $\rho_{\bm Q_1}(\bm r)$ and $\rho_{\bm Q_2}(\bm r)$ are non-vanishing this state is a bi-directional crystal. However, if $\rho_{\bm Q_1}(\bm r)\neq 0$ and $\rho_{\bm Q_2}(\bm r)=0$ (or vice-versa), this state is a unidirectional CDW (or stripe).  In the first case there are two spontaneously broken translations and the symmetry  is $U(1) \times U(1)$, while in the latter there is only one broken translation and the symmetry is just $U(1)$.
However, unlike superconductors, the stripe (and the CDW) are particle-hole condensates they are \textit{charge-neutral} and do not couple to the electromagnetic field. In this case the $U(1)$ symmetry represents continuous translations in one direction (given by $\bm Q$) and the corresponding uniform shift of the phase of the order parameter $\rho_{\bm Q}$.

The order parameter of a two-dimensional  nematic is any $2 \times 2$  traceless tensor $\mathcal{Q}$
\begin{equation}
\mathcal{Q}=
\begin{pmatrix}
\mathcal{Q}_{11} & \mathcal{Q}_{12}\\
\mathcal{Q}_{12} & -\mathcal{Q}_{12}
\end{pmatrix}
\label{eq:traceless-symmetric}
\end{equation}
Here $1$ and $2$ represent two orthogonal directions, say $x$ and $y$. A tensor of this type transforms as a quadrupole and   is invariant under rotations by $\pi$. One can also use the two non-vanishing components of  the tensor $\mathcal{Q}$ to for a \textit{director}: a vector with an orientation but not a direction. There are many choices for this tensor. Phenomenologically one can use any physical observable which transforms properly under rotations  such as the resistivity tensor
\begin{equation}
\rho=
\begin{pmatrix}
\rho_{xx} & \rho_{xy}\\
\rho_{xy} & \rho_{yy}
\end{pmatrix}
\label{eq:resistivity-tensor}
\end{equation}
In this case one can take (up to a normalization) $\mathcal{Q}_{11} \propto \rho_{xx}-\rho_{yy}$ and $\mathcal{Q}_{12} \propto \rho_{xy}$, which is commonly done in transport experiments. 

The nematic phase can also be regarded as an example of what has come to be known as \textit{vestigial order} \cite{Nie-2013}, that is as the remnant of a phase that breaks translation invariance such as a spin or charge density wave. In this case the nematic order parameter can be taken to be 
\begin{equation}
\mathcal{N}=|\rho_{\bm Q_1}|^2-|\rho_{\bm Q_2}|^2
\label{eq:Ising-nematic}
\end{equation}
 in a state in which $\langle \rho_{\bm Q_1}\rangle=\langle \rho_{\bm Q_2}\rangle =0$. In this case the nematic order parameter reflects the symmetry breaking of the point group $C_4 \to C_2$ and is called an Ising nematic. The same analysis holds for a spin-density wave state which has been used  to characterize  the nematic phase of iron superconductors. 

The electronic liquid crystals differ from the classical counterparts in several important ways. One is that since they are made of electrons, the smectic phase is actually a conductor in one direction and an insulator in the other. In other words it is a smectic metal.  On the other hand the nematic phase is an anisotropic metal. Also, while in classical liquid crystals the nematic is a phase in which rod-like nematic molecules (`nematogens') form a translation invariant anisotropic phase, in the present case this anisotropic phase is self-organized.

In addition,  the smectic metal can be regarded in a loose sense as as a array of one dimensional conductors, an array of Luttinger liquids, their interactions (in the form of Josephson couplings)  can result in an anisotropic superconducting state. This is a novel mechanism which is not BCS type. We explored the consequence of this physics in subsequent work with my collaborators \cite{Emery-2000,Granath-2001} which corroborated and extended the arguments of Ref.\cite{Kivelson-1998} This body of ideas showed that   that the emergence of charge order is not only not necessarily in conflict with superconductivity but it can also lead to a superconducting state whose critical temperature is not exponentially small as in the weak coupling BCS mechanism \cite{Arrigoni-2004,Kivelson-2007,Fradkin-2015}.

While stripe phases (``smectic'') were known from the work of Tranquada and coworkers, the nematic phase was a conjecture of our work of Ref.\cite{Kivelson-1998}. Remarkably the first (and to date) the most spectacular experimental confirmation of the existence of electronic nematic phases was found in beautiful experiments in two-dimensional electron fluids in magnetic fields \cite{Lilly-1999,Du-1999,Schreiber-2020}, and extended the application of the concept of electronic liquid crystal phases to these most strongly correlated systems \cite{Fradkin-1998}. By now, electronic nematic phases have also been seen experimentally  in several of systems such as the bilayer ruthenate {\SROtwo} \cite{Borzi-2007} (in a window of magnetic fields), in iron-superconductors \cite{Fernandes-2019}, and in the copper-oxide superconductors \cite{Hinkov-2008,Chang-2011} (for review see Ref. \cite{Fradkin-2010} ).

We should note that analogs of the electronic liquid crystal phases have been argued to play a role in the structure of the crusts of neutron stars \cite{Ravenhall-1983,Lorenz-1993}. In that context they are called ``pasta phases''.

\section{Competing Orders}
\label{sec:competing}

\begin{figure}[hbt]
\center
\includegraphics[width=0.450\textwidth]{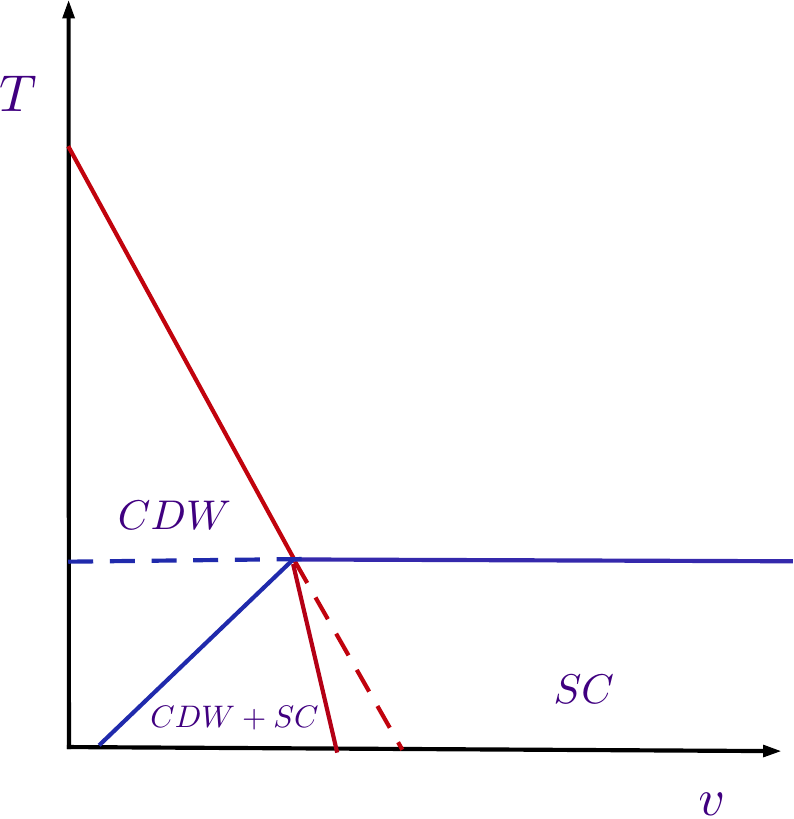}
\caption{Schematic phase diagram for competing  CDW and superconductivity (SC)  orders. Here $T$ is temperature and $v$ is the  coupling constant between the CDW and SC orders of Eq.\eqref{eq:competing}. See text for details.}
\label{fig:competing}
\end{figure}   

The existence of several types of orders (seemingly unrelated to each other) raises the problem that they may possibly compete with each other. However, this view cannot explain the generic occurrence of these different orders in many materials and over a  wide range  of parameters, such as doping, pressure, etc. To see the point we will consider a theory with two different orders. Specifically we will  consider the case represented by a complex field $\psi(\bm r)$ and $\eta(\bm r)$, but this is easily generalizable. In this case the free energy has a global $U(1) \times U(1)$ symmetry. We can think of $\psi$ as representing superconductivity and $\eta$ representing an incommensurate CDW. The free energy density is
\begin{align}
\mathcal{F}=&
\kappa_{\rm SC} |\bm \nabla \psi|^2+(T-T^c_{\rm{SC}}) |\psi|^2+u_{\rm SC} |\psi|^4 \nonumber \\
&+\kappa_{\rm CDW} |\bm \nabla \eta|^2+(T-T^c_{\rm{CDW}})|\eta|^2+u_{\rm CDW} |\eta|^4+ v  |\psi|^2|\eta|^2
\label{eq:competing}
\end{align}
The right hand side of the first line is the free energy density of the superconductor and the second line is the free energy density of the CDW. Here $T^c_{\rm SC}$ and $T^c_{\rm CDW}$ are the superconductor and CDW critical temperatures, $\kappa_{\rm SC}>0$ and $\kappa_{\rm CDW}>0$ are their stiffnesses, and $u_{\rm SC}>0$, $u_{\rm CDW>0}$ and $v$ are coupling constants. For $v>0$ the two orders compete and for $v<0$ they coexist. However, in order for the two order parameters to be comparable in strength to each other one will have to fine tune the parameters of the free energy of Eq.\eqref{eq:competing} to be very close to a multi-critical point at which both $\psi$ and $\eta$ will be critical simultaneously \cite{Kosterlitz-1976}. A typical phase diagram is shown in Fig.\ref{fig:competing}. As shown in Fig.\ref{fig:competing}  this scenario does not exhibit a ``superconducting dome'' which is characteristic of strongly correlated materials.

This example represents the qualitative behavior of the dichalcogenide NbSe$_2$ in which case the CDW is the large order that sets in at high temperatures and SC only sets in at lower temperatures and in that case $T^c_{\rm SC} <T^c_{\rm CDW}$.  The microscopic origin is that at high temperatures  NbSe$_2$ is a good metal with well defined quasiparticles and the CDW and superconductivity compete.
Here CDW gaps out a large fraction of the Fermi surface and superconductivity happens on the remaining piece. This simple picture is the standard explanation for the disparity in the magnitudes of the critical temperatures. The only way that they can be comparable is to make the attractive interaction strong enough.

We should also note that since the CDW breaks translation invariance, with an ordering wave vector $\bm Q$, the superconducting state should necessarily acquire a modulated component with the same ordering wave vector as the CDW. As it stands this physics is not reflected in the free energy of Eq.\eqref{eq:competing}. In Section \ref{sec:topo-PDW} we will come back to this question. This effect has been seen in   STM experiments with a superconducting tip in NbSe$_2$ \cite{Liu-2021}.

There is another interesting effect of systems with coupled order parameters of the form of Eq.\eqref{eq:competing} is the appearance of ``halos'' of a competing order parameter in the vicinity of a topological defect of the dominant order. For example  consider the case of the  Landau theory of Eq.eqref{eq:competing} in the regime in which the orders compete, $v>0$ and in the temperature range  $T^c_{\rm CDW}<T<T^c_{\rm SC}$  the superconducting order parameter has an expectation value $\langle \psi \rangle \neq 0$ and and the CDW order parameter $\langle \phi \rangle=0$. In the presence of a magnetic field, in the mixed phase of the superconducting state (assuming that it is type II), the superconducting order parameter is suppressed near the core of the vortex. For a range of parameters of the Landau free energy, even for $T>T^c_{\rm CDW}$, the competing CDW order parameter is induced inside a ``halo'' with a length scale longer than the size of the superconducting core \cite{Kivelson-2002}. I will comment more on ``halos'' in Section \ref{sec:topo-PDW}.

This halo effect has been seen in neutron scattering experiments in the mixed state of {\LSCO} in a field $B\sim 8T$ \cite{Lake-2002} where the competing order is an SDW (a spin stripe), and in STM experiments in the superconducting state of {\BSCCO} in a magnetic field \cite{Hoffman-2002} where the competing order is a CDW (a charge stripe). Naturally, the same idea works with other order parameters, e.g. with the nematic order instead of  a CDW \cite{Chowdhury-2011}. The reverse effect, i.e. a defect of charge order inducing superconductivity, has also been discussed: superconductivity being nucleated at a domain wall of a period 2 CDW (also works for nematic order) for instance in Cu intercalated dichalcogenide $1T$-TiSe$_2$ \cite{Yan-2016} and in the pnictide {\BSNA} \cite{Lee-2021}. It is possible to generalize this notion for other periods, where the CDW order parameter is complex although the effects are weaker than for period 2 since it requires a derivative coupling \cite{Castro_Neto-2019}. These halo effects are more pronounced close to a phase transition of the subdominant order.

\section{Pair Density Waves and Intertwined Orders}
\label{sec:intertwined}

Away from a multi-critical point, the magnitudes of coupled order parameters described by a Landau theory of competing orders of the form of Eq.\eqref{eq:competing} rapidly become very different from each other. While this is perfectly correct prediction of Landau theory (and its generalizations), it  does not describe what is seen experimentally in strongly correlated materials such as the copper oxide superconductors (and many others). In these systems different order parameters have comparable critical temperatures over a broad range of materials parameters without the need to fine tuning to a quantum (or thermal) phase transition. This observation  suggests that many of the observed phases in these materials may arise from common origin and originate from the same microscopic physics. With my collaborators we  have called this behavior as \textit{Intertwined Orders} \cite{Fradkin-2015,Berg-2009}.

\subsection{Intertwined Orders in {\LBCO}}
\label{sec:LBCO}

\begin{figure}[hbt]
\center
\includegraphics[width=0.550\textwidth]{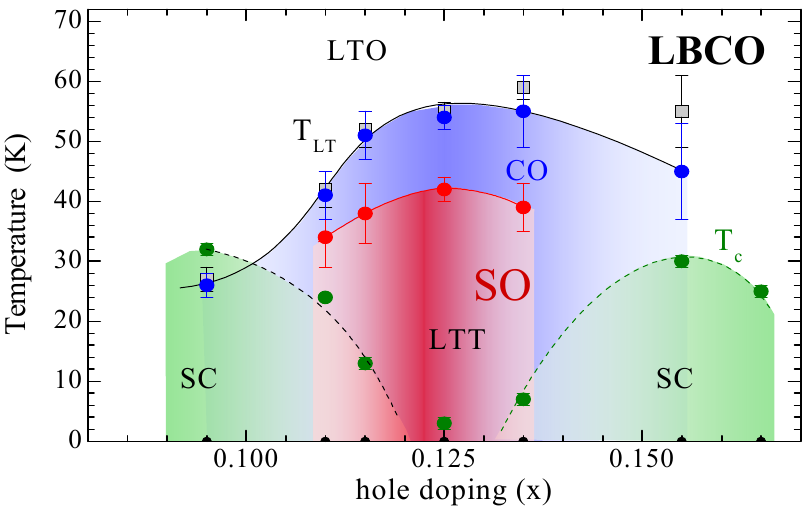}
\caption{Experimental temperature vs doping phase diagram of {\LBCO}. CO: charge order; SO: spin order; SC: superconductivity; LTO: low temperature orthorhombic phase; LTT: low temperature tetragonal phase. (from the work of M. H\"ucker and coworkers \cite{Hucker-2011}. )}
\label{fig:LBCO}
\end{figure}

The physics of intertwined orders is apparent  {\LBCO}. As we noted before this material is where high temperature superconductivity was discovered by Bednorz and M\"uller in 1986 \cite{Bednorz-1986}. However it was not subsequently studied in detail for two reasons. One was the discovery of {\YBCO} which has a much larger critical temperature. The other is that at doping $x\sim 1/8$ the critical temperature of {\LBCO} is strongly suppressed from $T_c \sim 35$K down to $T_c\sim 4$K. Because of this the phase diagram of {\LBCO}, shown in Fig.\ref{fig:LBCO} has two domes instead of one and it resembles the back of a Bactrian camel.

{\LBCO} has a complex structure. At relatively high temperatures the crystal structure is in the low temperature orthorhombic (LTO) phase and there is a doping-dependent structural transition, with a broad maximum for $x \sim 1/8$ at about $52$ K, to the low temperature tetragonal (LTT) phase. In the LTT phase the copper oxide planes form pairs in which each plane is orthorhombic but rotated by $90^\circ$ from the other plane in the pair. In the LTT phase of {\LBCO} there is static charge stripe order (denoted by CO in Fig.\ref{fig:LBCO}) which is unidirectional in each plane, with periodicity of $4$ lattice constants, and rotated by $90^\circ$ between planes \cite{Fujita-2004,Abbamonte-2005}.   In other materials of the Lanthanum series of the cuprates, e.g. in {\LESCO} , the LTO-LTT structural transition happens at much higher temperatures than the charge ordering transition (see Ref.\cite{Lee-2022}.) In the LTO phase the charge stripe order is ``fluctuating'' (in the sense of Ref.\cite{Kivelson-2003}) seen only in the inelastic channel at low frequencies and with a doping-dependent ordering wave vector \cite{Miao-2019}.
At $x \sim 1/8$ a cascade of phase transitions are seen below $52$K where static charge stripe order is seen. Then at $\sim 42$K there is a transition a phase with static stripe  unidirectional spin order with ordering wave vector $\bm Q_{\rm SDW}=2\bm Q_{\rm CDW}$. Static charge and spin orders are  commensurate with each other.

 A number of remarkable transport anomalies are seen  below the static spin ordering temperature \cite{Li-2007,Tranquada-2008}. Already in the LTO phase $\rho_c$, the c-axis resistivity (normal to the Cu-O planes) and the in-plane resistivity $\rho_{ab}$, differ by a factor of $\sim 2 \times 10^3$, and begin to gradually depart even more in the LTT phase. Below the spin ordering transition $\rho_c$ continues to increase with lowering temperatures while $\rho_{ab}$ begins to decrease rapidly. After a broad fluctuation regime $\rho_{ab}$ becomes unmeasurably small while $\rho_c$ had continued to increase. In other words, the Cu-O planes became superconducting while the c-axis transport is highly insulating! This anomalous behavior, in which the Cu-O planes decouple from each other and the c-axis phase coherence (or, equivalently, the c-axis Josephson effect) is absent. A similar effect had been seen before  by Tajima et al \cite{Tajima-2001} in {\LNSCO}. The full 3D resistive transition, where $\rho_c \to 0$, is reached at $T_{\rm resistive} \sim 10$K. However, even though below $T_{\rm resistive}$ $\rho_c=\rho_{ab}=0$, no Meissner effect is seen until $T_{\rm Meissner} \sim 4$ K where the Meissner effect is seen. This is the actual ``bulk''  critical temperature of this superconductor. This large separation between the resistive transition and the actual superconducting transition is also anomalous.

The remarkable  \textit{dynamical layer decoupling} means that the interlayer Josephson effect has been suppressed. Away from $x=1/8$ the charge order weakens,  the superconducting  $T_c$ is large and  the dynamical layer decoupling is absent. However, even a modest perpendicular magnetic field restores  the layer decoupling reappears \cite{Hucker-2011,Hucker-2013}. This effect, the suppression of the c-axis Josephson resonance,  has been seen in many Lanthanum materials and was studied in detail in {\LSCO} by Schafgans and collaborators \cite{Schafgans-2010}. 

In 2007 we proposed that the dynamical layer decoupling can be explained as evidence of a new superconducting state that we called a \textit{Pair Density Wave} (PDW) which competes with the uniform-d-wave superconducting state \cite{Berg-2007}. The PDW is a superconducting state in which the order parameter oscillates in space and, hence, in this state the Cooper pairs have finite momentum. In this  sense the PDW order parameter is the same as that of the Larkin-Ovchinnikov state \cite{Larkin-1965} (LO), proposed long ago (but to this date not observed). However, the LO state requires an external magnetic field to be present and the modulation of the state is determined by the field strength. Thus, the LO breaks time-reversal invariance, whereas the PDW is self-organized and is time-reversal invariant. In addition, since it is controlled by the strength of the magnetic field, the typical period of the LO state is necessarily very long. Instead,  the PDW state has a periodicity of a few lattice spacings.

To this date the layer decoupling effect has clearly only been seen  in the Lanthanum series of the copper oxide superconductors.  We will see below that the PDW state in a system with the symmetry of the LTT crystal  structure  yields a natural explanation of the suppression of the interlayer Josephson effect. However, this does not mean that it is restricted to materials with that crystal structure. We will see below that there is evidence for the PDW as a competing state of the d-wave superconducting state, or as a subdominant order, in other materials.

\subsection{The Pair-Density Wave State}
\label{sec:PDW}

\begin{figure}[hbt]
\center
\includegraphics[width=0.450\textwidth]{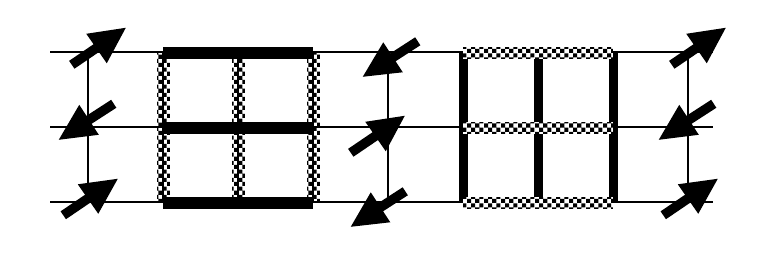}
\caption{Qualitative picture of the PDW state proposed for {\LBCO} at the 1/8 anomaly: spin, charge and superconducting orders are closely intertwined with each other is this period 4 PDW in a locally d-wave superconducting state (from the work of E. Berg and coworkers, Ref. \cite{Berg-2007}). See text for details.}
\label{fig:intertwined-stripe}
\end{figure}

With E. Berg and several other collaborators we proposed an explanation of the anomalies observed in {\LBCO} at the 1/8 anomaly as being due to a state in which charge, spin and superconducting d-wave states are closely intertwined with each other \cite{Berg-2007}. In our proposal the superconducting state is an inhomogeneous state with period 4 lattice spacings which on local stripe-like regions is a d-wave state which are rotated by $\pi/2$ and separated by domain walls where there is a charge accumulation and local antiferromagnetic order, see Fig.\ref{fig:intertwined-stripe}. In this state the antiferromagnetic order has period 4, the same as the superconducting order, while the charge order has period 2. Thus, these orders are commensurate with each other. 

\begin{figure}[hbt]
\center
\includegraphics[width=0.450\textwidth]{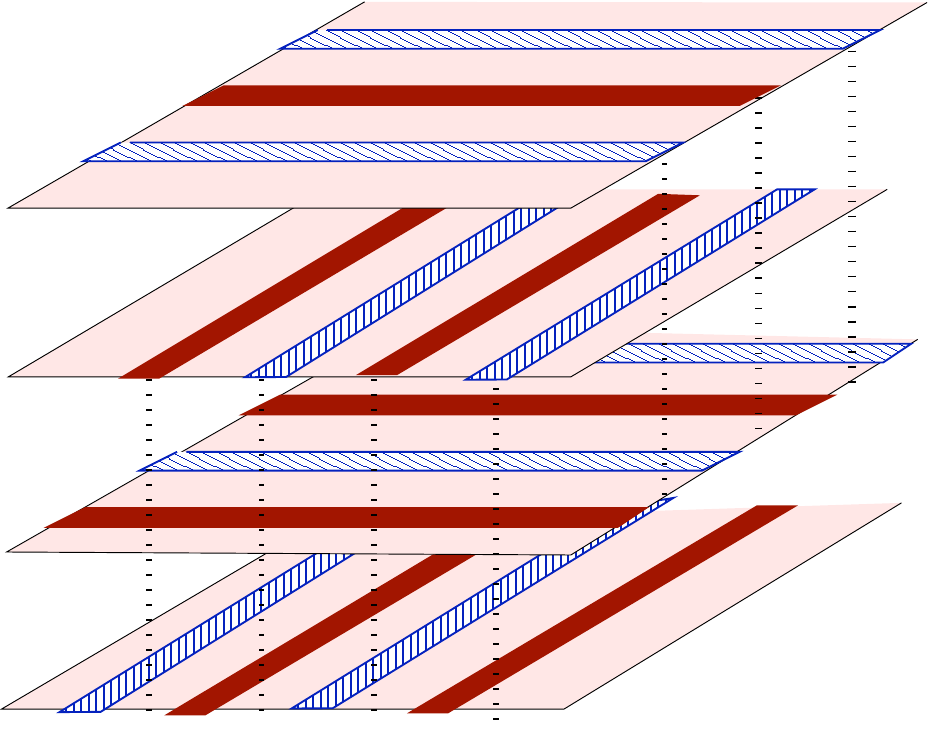}
\caption{Qualitative picture of the PDW on the LTT crystal structure of {\LBCO} (from the work of E. Berg and coworkers, Ref. \cite{Berg-2007}). See text for details.)}
\label{fig:stacked}
\end{figure}

On the other hand, the crystal structure of {\LBCO} is LTT meaning that consecutive copper oxide planes are orthorhombic and $\pi/2$ rotated from each other. Coulomb repulsion force copper oxide planes with the same orthorhombic symmetry to be shifted from each other by half a period of the charge order. The net result is that the unit cell must have four layers along the c-axis and be enormously elongated in that direction and the planes are stacked in the pattern  shown in fig.\ref{fig:stacked}.

How does this state explain the observed anomalies  and, in particular, the dynamical layer decoupling? First the rotation of consecutive planes suppresses the inter-layer Josephson effect (pair-tunneling) since locally the superconducting orders are phase shifted by $\pi$. Notice that it does not suppress the \textit{double Josephson effect}, i.e. tunneling of \textit{quartets}. In addition the spins are also rotated by $\pi/2$ (dues to spin-orbit effects). Finally, Coulomb repulsion forces the unit cell to have period 4 along the c-axis. Thus, the planes with an allowed Josephson coupling are extremely far from each other and the effective coupling is essentially negligible. The result is that the Josephson effect is frustrated and the planes are effectively decoupled. Away from 1/8 doping, defects, such as local charge disorder, and dislocations of the charge order lead to a small but finite Josephson coupling between nearby planes and favor the uniform superconducting order and the suppression of the dynamical layer decoupling.

Clearly, this state relies on strong correlation physics. Indeed, variational Monte Carlo simulations of the $t-J$ model on a square lattice by Himeda, Kato and Ogata \cite{Himeda-2002} and  later iPEPS simulations by Corboz, Rice and Troyer \cite{Corboz-2014} have found that the uniform d-wave SC, the striped SC and and the PDW are are extremely close in energy, with the uniform d-wave SC state being slightly preferred. These simulations suggest that the ordering between these ground states can be affected by even small changes in the models. The natural conclusion is that at the local level strong correlation physics favors a d-wave SC but whether it is uniform or not depends on details and that the PDW and the uniform state are in close competition.

\subsection{Landau Theory Picture of the PDW State}
\label{sec:landau-PDW}

We now turn to the order parameter description of the PDW state. The PDW is a superconducting state which, in addition of breaking spontaneously the phase symmetry, breaks translation and rotation invariance. In this sense, it is a superconducting electronic liquid crystal state.  Let $\Delta (\bm r)$ be the local superconducting amplitude. In an inhomogeneous state we can expand the fluctuations of the local pairing amplitude in Fourier components
\begin{equation}
\Delta(\bm r)=\Delta_0(\bm r)+\sum_{i=1,2}\Big[\Delta_{\bm Q_i}(\bm r) \exp(i \bm Q_i \cdot \bm r)+\Delta_{-\bm Q_i}(\bm r) \exp(-i \bm Q_i \cdot \bm r)\Big]+\ldots
\label{eq:FT-Delta}
\end{equation}
where $\Delta_0(\bm r)$ is the local amplitude of the uniform order parameter and $\Delta_{\pm \bm Q_i}(\bm r)$ are the local amplitudes of the two PDW \textit{order parameters} with ordering wave vectors $\bm Q_i$ (with $i=1,2$, the two orthogonal directions of teh square lattice), and the ellipsis represents higher harmonics of the Fourier expansion. For simplicity, I am considering only states with unidirectional order (``stripes''), represented by a single ordering wave vector, say $\bm Q_1\equiv \bm Q$, and no modulation along direction 2. States with multiple ordering wave vectors have been considered in other systems such as the kagome superconducting materials \cite{ Li-2023}, in UTe$_2$ \cite{Aishwarya-2023,Gu-2023}, and in the iron pnictide {\EuRbFeAs} \cite{Zhao-2023}. Here we are assuming that the superconductivity has spin singlet pairing. The extension to a spin triplet PDW was considered in Ref.\cite{Aishwarya-2023}.

The representation of Eq.\eqref{eq:FT-Delta} describes different phases: 
a \textit{uniform superconductor} is  $\langle \Delta_0 \rangle \neq 0$ and $\langle \Delta_{\bm Q}\rangle = \langle \Delta_{-\bm Q} \rangle =0$, 
b)  a \textit{PDW} if $\langle \Delta_0 \rangle = 0$ and $\langle \Delta_{\rm Q} \rangle= \langle \Delta_{-\rm Q}\rangle \neq 0$, 
c)  an \textit{inversion-breaking superconductor} if $\langle \Delta_0 \rangle = 0$, $\langle \Delta_{\rm Q} \rangle \neq 0$ and  $\langle \Delta_{-\rm Q}\rangle = 0$,
and d) a \textit{striped superconductor} if $\langle \Delta_0 \rangle \neq 0$, $\langle \Delta_{\bm Q} \rangle \neq 0$ and $\langle \Delta_{-\bm Q} \rangle \neq 0$. Each one of these states has distinct symmetry properties. The inversion-breaking superconductor was predicted by Fulde and Ferrell \cite{Fulde-1964} (FF) as arising in the presence of a magnetic field much in the same way as the state proposed by Larkin and Ovchinnikov. The LO and FF states are commonly referred together as the FFLO states (even though they have different symmetry) and the there is a robust literature on them as magnetic-driven superconducting states \cite{Casalbuoni-2004} or as arising from a Lifshitz transition at which the superfluid stiffness becomes negative. For these reasons the FFLO states have very long wavelengths, typically much longer than the lattice spacing of a material.  Instead, here we are interested in these translation symmetry breaking superconducting states as arising in the absence of an external magnetic field and having periodicities of a few lattice constants. On the other hand, with the exception of the Lanthanum family of the coper oxide superconductors, most states normally referred to as  a PDW in the literature are actually striped superconductors. We will see below why this distinction matters.

%{\BaFeAs122} \cite{Schmidt-2019}

Here we will focus on a unidirectional PDW state first and later discuss its interplay with the uniform state. The multidimensional case is similar but with a richer set of phases. This has been discussed in several references  \cite{Agterberg-2008,Agterberg-2015,Berg-2009,Berg-2009a,Fradkin-2015,Agterberg-2020}. In addition, the CDW order parameters $\rho_{\bm K}$, SDW order parameters $\bm S_{\bm P}$ and the nematic order parameter $\mathcal{N}$   can be regarded as being induced (``daughter'') orders (or vestigial orders \cite{Nie-2013}), defined in terms of composite order parameters,  or being ``preexisting'' and separately defined. Below we  discuss this distinction.

Since the unidirectional PDW superconductor has \textit{two} complex order parameters, which we  denoted by $\Delta_{\pm \bm Q}(\bm r)$ \cite{Berg-2009,Berg-2009a},  this state is described  by two amplitudes $|\Delta_{\pm \bm Q}(\bm r)|$ and two phase fields that we will denote by $\theta_{\pm \bm Q}(\bm r)$, each defined mod $2\pi$. The free energy density of the PDW state  has a $U(1) \times U(1)$ global symmetry under which each phase field is shifted by constant (but different)  phases,  $\alpha_\pm$ (each defined mod $2\pi$), so that the free energy is invariant under the global change $\theta_{\pm \bm Q}(\bm r) \mapsto \theta_{\pm \bm Q}(\bm r)+\alpha_{\pm}$. In addition to these global symmetries the both PDW order parameters couple in the same way to the electromagnetic gauge transformation, as charge 2 bosonic field.  These symmetries dictate the form of the Landau free energy density.

A consequence of the PDW being a condensate of Cooper pairs with finite momentum is that the Fermi surface is not fully gapped. Instead, it consists of ``pockets'' with a Bogoliubov Fermi surface with one side being predominantly electron-like and the other hole-like \cite{Berg-2009,Soto-Garrido-2015,Agterberg-2018}.

The free energy density of the unidirectional PDW is a generalization of Eq.\eqref{eq:GL} 
\begin{align}
\mathcal{F}_{\rm PDW}=&\kappa_{\rm PDW} \left|\left(\bm \nabla +i \, \frac{2e}{\hbar c} {\bm A}(\bm x)\right)\Delta_{\bm Q} (\bm x) \right|^2+(T-T_c^{\rm PDW}) |\Delta_{\bm Q}(\bm x)|^2+\frac{u_{\rm PDW}}{2} |\Delta_{\bm Q}(\bm x)|^4 \nonumber\\
+&\kappa_{\rm PDW} \left|\left(\bm \nabla +i \, \frac{2e}{\hbar c} {\bm A}(\bm x)\right)\Delta_{-\bm Q} (\bm x) \right|^2+(T-T_c^{\rm PDW}) |\Delta_{-\bm Q}(\bm x)|^2+\frac{u_{\rm PDW}}{2} |\Delta_{-\bm Q}(\bm x)|^4 \nonumber\\
+&v_{\rm PDW} |\Delta_{\bm Q}|^2 |\Delta_{-\bm Q}|^2
\label{eq:GL-PDW}
\end{align}
For $v_{\rm PDW}>0$ either $\Delta_{\bm Q}\neq 0$ and $\Delta_{-\bm Q}=0$ (or viceversa) this is an FF state. For $v_{\rm PDW}<0$, 
$\Delta_{\bm Q}=\Delta_{-\bm Q}\neq 0$ and this is the PDW state. For simplicity I ignored  that the stiffness of a unidirectional PDW $\kappa_{\rm PDW}$ is anisotropic .

%In Eq.\eqref{eq:GL-PDW} we have assumed that inversion symmetry is exact which requires that the stiffnesses for $\Delta_{\bm Q}$ and $\Delta_{-\bm Q}$ must be the same. If inversion symmetry is explicitly broken the stiffnesses will generally be different. However in this case the critical temperatures will generally be different and there will be an additional FF phase in the phase diagram. Here we will assume that inversion symmetry is exact.

The PDW order parameters $\Delta_{\pm \bm Q}$ and the  CDW order parameter $\rho_{\bm K}$ can couple to each other through the standard  biquadratic interaction, $w |\Delta_{\pm \bm Q}|^2 |\rho_{\bm K}|^2$, which is always allowed, as in Eq.\eqref{eq:competing}, and describes the competition/coexistence between CDW and PDW orders.  However, a trilinear interaction of the form 
\begin{equation}
\mathcal{F}_{\rm trilinear}=\gamma \, \rho_{\bm K}^* \Delta_{-\bm Q}^*\Delta_{\bm Q}+\textrm{c.c.}
\label{eq:trilinear}
\end{equation}
is allowed provided  the ordering wave vectors are related by the mutual commensurability condition $\bm K=2 \bm Q$. This is required by translation invariance. Notice that for $\gamma >0$ there will be a relative phase  of $\pi/2$ between $\Delta_{\bm Q}$ and $\Delta_{-\bm Q}$ which implies that in this case time-reversal symmetry will be broken spontaneously. This phase difference is absent if $\gamma<0$.

Eq.\eqref{eq:trilinear} has two interpretations. One is that if the CDW is an independent order parameter (i.e. preexisting) it \textit{can} trigger the formation of a PDW which would require a phase transition at which the gauge symmetry of both PDW order parameters is spontaneously broken. This can only happen if the amplitude of the CDW is large enough or the coupling constant $\gamma$ is strong enough. This is a reasonable scenario for {\LBCO} given the high critical temperature for the charge order transition.
Experiments also show that the strength of the CDW order weakens significantly away for $x=1/8$ with a concomitant  increase of the critical temperature of the uniform superconducting order is seen \cite{Hucker-2011}. However, this interpretation has to be considered with some reservation given the inhomogeneity of the material. 

The other interpretation of Eq.\eqref{eq:trilinear}  is that, even in the absence of a preexisting CDW, in the PDW phase a CDW with wave vector $\bm K=2\bm Q$ will necessarily be present. In this sense, we can regard the PDW as having a ``daughter'' (or induced) composite order parameter, 
\begin{equation}
\rho_{\bm K} \sim \Delta^*_{\bm -Q}\Delta_{\bm Q}
\label{eq:daughter-CDW}
\end{equation}
which is gauge-invariant and has wave vector $\bm K=2\bm Q$. 

The same line of reasoning implies that the PDW has an induced uniform composite order parameter
\begin{equation}
\Delta_{4e}\sim \Delta_{\bm Q}\Delta_{-\bm Q}
\label{eq:daughter-4e}
\end{equation}
which has zero momentum and transforms under gauge transformations as a charge $4e$ complex field, i.e. a  \textit{quartet condensate}. We will see shortly that the existence of the quartet condensate has important consequences. 

We should note that quartet condensates were considered long ago in Nuclear Physics. However, in the absence of  some symmetry, quartet  condensates can decay into Cooper pairs. In fact, even in a conventional superconductor, once the pair amplitude has an expectation value, the quartet amplitude will also have an expectation value which will be  exceedingly small. For instance, in a Josephson junction with a phase difference $\Delta \phi$, the (Josephson) current is $\sim J \sin(\Delta \phi)$. However for $J$ large enough there will be a significant amplitude for the harmonic $J_4 \sin (2\Delta\phi)$ (representing the tunneling of pairs of pairs). Typically $J_4 \sim J^2$ and since $J$ is exponentially small, $J_4$ is even smaller. 

On the other hand, in the PDW the Cooper pairs have \textit{finite momentum} whereas the quartet amplitude $\Delta_4$ has \textit{zero momentum}. Thus, provided momentum conservation is a good symmetry the quartets cannot decay into Cooper pairs. In fact, although by symmetry Josephson coupling between copper oxide planes of {\LBCO} is suppressed by symmetry, tunneling of quartets (i.e. pairs of pairs) is not.

Let us now consider what changes if there is also a uniform component $\Delta_0$. The free energy now will be modified in several ways. One is that there will be the usual conventional terms  for $\Delta_0$ have the form of Eq.\eqref{eq:GL} albeit with different parameters than the terms associated with the PDW. Bearing in mind the case of {\LBCO}, I will assume that the uniform state only appears at lower temperatures and $T_c^0<T_c^{\rm PDW}$. In the case of {\LBCO} at $x=1/8$ doping the Meissner state has a $T_c \sim 4$K. In this phase one expects to see d-wave superconducting order and a conventional Josephson coupling. Hamilton and collaborators \cite{Hamilton-2018} reported that phase sensitive measurements checked that {\LBCO}  is, as expected, a d-wave superconductor. Furthermore, in-plane Josephson tunneling from Nb experiments revealed that, in addition to the conventional $\sin (\Delta \phi)$ Josephson current in {\LBCO} at $x=1/8$ there is also a smaller $\sin(2\Delta \phi)$ component. This experiment also showed that at $x=1/8$ doping while the $\sin (\Delta \phi)$ component decreases with increasing temperature, the  $\sin(2\Delta \phi)$ increases with increasing temperature. This effect is absent in doping regimes where the superconducting $T_c$ is highest.

Of particular interest are the terms that can couple the order parameters $\Delta_0$ and $\Delta_{\pm \bm Q}$. As usual, there is always a biquadratic interaction of the form $\gamma' |\Delta_0|^2 |\Delta_{\pm \bm Q}|^2$. As before, if $\gamma'>0$ the two orders compete and if one is present the other is absent, and if $\gamma'<0$ the two orders coexist. 

On the other hand a term involving the  coupling of the \textit{phase fields} of $\Delta_0$ and $\Delta_{\pm \bm Q}$  of the form
\begin{equation}
\mathcal{F}_{\rm locking} = \tilde \gamma \, \Delta_0^2 \Delta_{\bm Q}^* \Delta_{-\bm Q}^*+\textrm{c.c.}
\label{eq:phase-locking}
\end{equation}
can lock the phase fields of the PDW and the uniform order. Here $\tilde \gamma$ is a coupling constant. 
This term implies that the phase field $\theta_0$ of $\Delta_0$ and  the average phase field of the PDW, $(\theta_{\bm Q}+\theta_{-\bm Q})/2$,  should be locked to each other, up to an $n\pi$ phase difference, where $n \in \mathbb{Z}$. For $\tilde \gamma<0$  the two orders coexist and there is a $\pi/2$ phase difference between the  order parameters which breaks time-reversal invariance spontaneously \cite{Wu-2025}. 

Another trilinear term in the free energy is possible in the  presence of the uniform order parameter 
\begin{equation}
\mathcal{F}_{\rm trilinear}'=\eta \, \rho_{\bm K} \Delta_0^* \Delta_{\bm Q}+{\rm c.c.}
\label{eq:1Q}
\end{equation}
where $\eta$ is a coupling constant. Translation invariance the requires that is  allowed only  if $\bm K=\bm Q$. 

The  trilinear term of Eq.\eqref{eq:1Q} has several implications. One is that if both CDW and uniform superconducting orders are present, $\langle \rho_{\bm K}\rangle\neq 0$ and $\Delta_0\neq 0$, then $\langle \Delta_{\bm Q}\rangle \neq 0$. This means that there will be a modulated component of the superconducting order parameter with the same ordering wave vector of the charge order. This modulated component was see in STM experiments with a superconducting tip in NbSe$_2$ \cite{Liu-2021}, in {\BSCCO} \cite{Hamidian-2015,Chen-2022}, and in the heavy-fermion superconductor UTe$_2$\cite{Gu-2023}. 

The other consequence of Eq.\eqref{eq:1Q} is that if $\langle \Delta_0 \rangle \neq 0$ and $\langle \Delta_{\bm Q} \rangle \neq 0$ then $\langle \rho_{\bm K}\rangle\neq 0$. Thus if both the uniform component and the PDW component are present then a CDW order is induced with same same ordering wave vector $\bm Q$ as the PDW, in addition to the $2\bm Q$ CDW order already present in the PDW state. Alternatively we can say that if $\langle \Delta_0 \rangle \neq 0$ and $\langle \Delta_{\bm Q} \rangle \neq 0$ then there will be an induced $\bm K=\bm Q$ component (usually referred to as the``$1\bm Q$'' component)  in the CDW order with $\rho_{\bm Q}\sim \Delta_0^* \Delta_{\bm Q}$. 

The prediction of this change in the pattern of the charge order is a prediction of the Landau theory and has been a focus of many studies in several materials, including {\YBCO} where it has not been seen in X-ray experiments (at least in one sample near optimal doping) \cite{Blackburn-2023}, in {\BSCCO} where it has been seen in the vortex halo in STM experiments in a magnetic field \cite{Edkins-2018}, and in Fe-Sr doped {\LBCO} at $x=1/8$ \cite{JS_Lee-2023} where it has been seen in the  Meissner phase of this material which has a enhanced $T_c \sim 10$K.

\subsection{Topological excitations of the PDW superconductor}
\label{sec:topo-PDW}

We will now discuss the topological textures of a PDW. here I will follow the treatment of Berg and coworkers \cite{Berg-2009b}. In  the case of a superconductor its topological textures are vortices while for CDWs they are dislocations \cite{Chaikin-1995}. However, the PDW has \textit{two} superconducting order parameters, the two complex fields that we called $\Delta_{\pm \bm Q}$. As a result the Landau-Ginzburg theory of Eq.\eqref{eq:GL-PDW} has a $U(1)\times U(1)$ global symmetry. Deep in the PDW phase the order parameters $\Delta_{\pm \bm Q}(\bm r)$, $\rho_{\bm K}(\bm r)$ and $\Delta_{4e}(\bm r)$ at long distances become
\begin{align}
\Delta_{\pm \bm Q}(\bm r) 
& \mapsto \Delta_{\rm PDW} \, \exp(i \theta_{\pm \bm Q}(\bm r))                      \equiv \Delta_{\rm PDW} \, \exp(i (\theta (\bm r)\pm\varphi(\bm r)))\\
\rho_{\bm K}(\bm r) 
& \mapsto \rho_{K}\, \exp(i (\theta_{\bm Q}(\bm r)-\theta_{-\bm Q}(\bm r))       \equiv \rho_K \, \exp(i2\varphi(\bm r))\\
\Delta_{4e}(\bm r) 
&\mapsto \Delta_{4e} \, \exp(i(\Delta_{\bm Q}(\bm r)+\Delta_{-\bm Q}(\bm r))) \equiv \Delta_{4e} \, \exp(i 2\theta(\bm r))
\label{eq:PDW-orders}
\end{align}
where $\Delta_{\rm PDW}$, $\rho_{K}$ and $\Delta_{4e}$ are finite positive real numbers which quantify the strength of each order. In what follows we assume that $\bm K=2\bm Q$ and CDW is an induced order commensurate with the PDW order, as defined in Eq.\eqref{eq:daughter-CDW}, and likewise with the charge $4e$ superconducting order defined in Eq.\eqref{eq:daughter-4e}.

Since the PDW order parameters $\Delta_{\pm \bm Q}(\bm r)$ must be separately single valued, the phase fields $\theta_{\pm \bm Q}$ must be invariant under the the periodic shifts
\begin{equation}
\theta(\bm r)_{\pm \bm Q}(\bm r) \mapsto \theta(\bm r)_{\pm \bm Q}(\bm r) +2\pi m_{\pm \bm Q}
\end{equation}
where $m_{\pm \bm Q} \in \mathbb{Z}$. The integers $m_{\pm \bm Q}$ are topological charges of the complex fields $\Delta_{\pm \bm Q}(\bm r)$.

Therefore the phase fields $\theta(\bm r)$ and $\phi(\bm r)$, defined in Eq.\eqref{eq:PDW-orders}, must obey the  conditions
\begin{align}
\theta(\bm r) &\mapsto \theta(\bm r) + \pi (m_{\bm Q}+m_{-\bm Q})\\
\varphi(\bm r) & \mapsto \varphi(\bm r)+\pi (m_{\bm Q}-m_{-\bm Q})
\label{eq:theta-phi}
\end{align}
these conditions  classify the topological textures of the PDW in terms of  a pair of topological charges $(q_s, q_c)$, which represent respectively vortices with topological charge $q_s=(m_{\bm Q}+m_{-\bm Q})/2$ and dislocations with topological charge $q_c=(m_{\bm Q}-m_{-\bm Q})/2$. With these definitions the phase fields $\theta(\bm r)$ 
and $\phi(\bm r)$ are defined mod $\pi$, consistent with the definitions of Eq.\eqref{eq:PDW-orders}.

\begin{figure}[hbt]
\center
\includegraphics[width=0.30\textwidth]{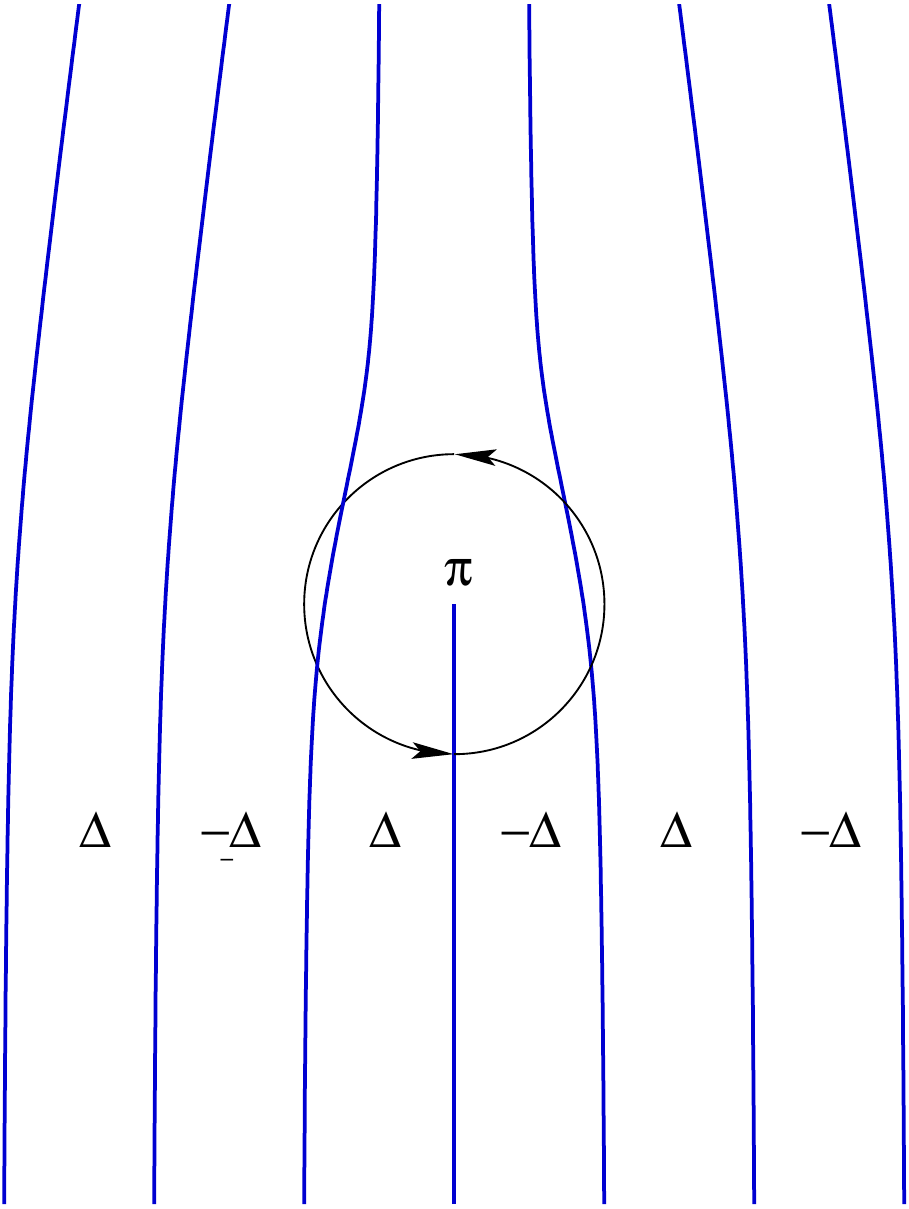}
\caption{The half-vortex. See text for details.)}
\label{fig:half}
\end{figure}  

\begin{figure}[hbt]
\center
\includegraphics[width=0.30\textwidth]{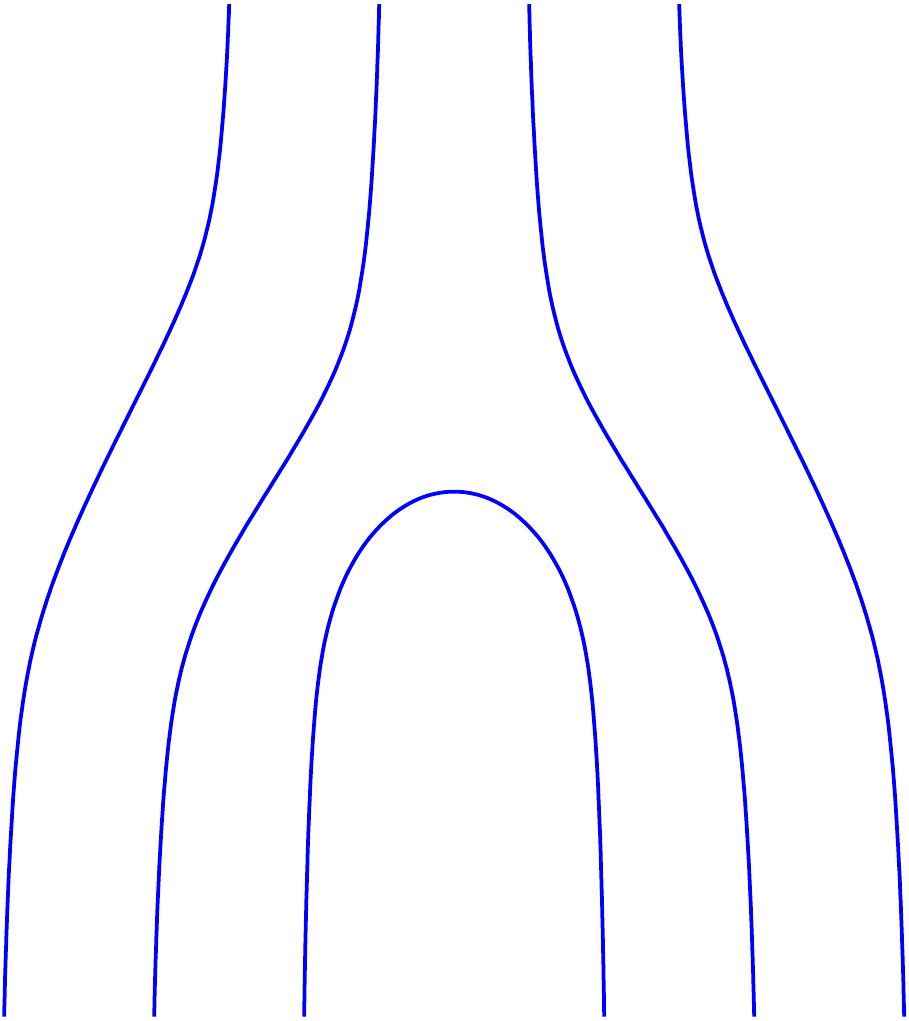}
\caption{The double dislocation). See text for details.)}
\label{fig:double}
\end{figure}

\begin{figure}[hbt]
\center
\includegraphics[width=0.70\textwidth]{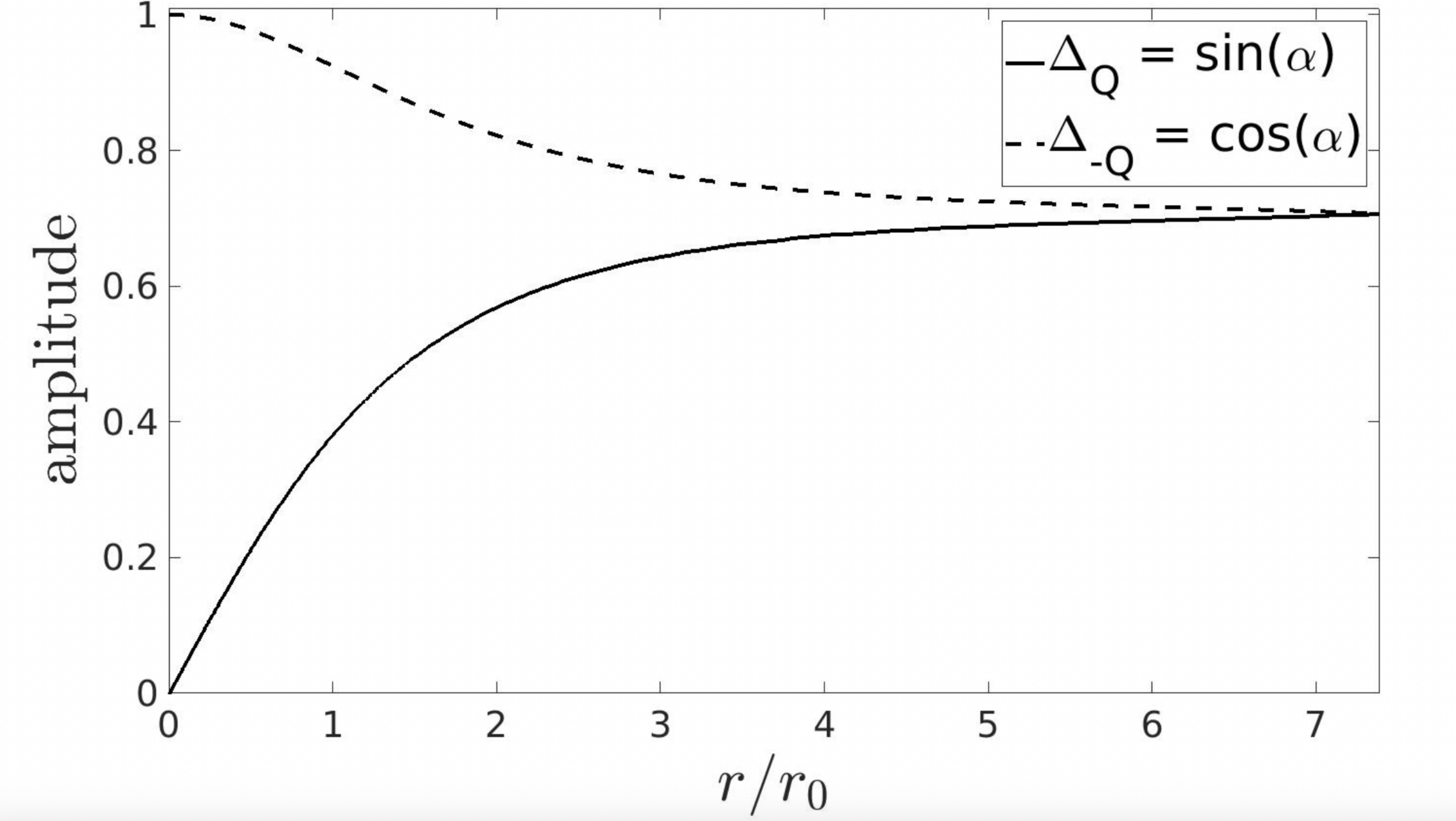}
\caption{Structure of the half vortex of a PDW: at the core of the half-vortex $\lim_{r \to 0} \Delta_{\bm Q}= 0$ but $\lim_{r\to 0} \Delta_{-\bm Q} \neq 0$ (and actually it increases from its value at long distances where $\lim_{r \to \infty} |\Delta_{\bm Q}|=\lim_{r \to \infty} |\Delta_{-\bm Q}|=\Delta_{\rm PDW}$). Hence, the core of the PDW vortex has an  ``FF halo'' which breaks inversion symmetry. In this halo the $2\bm Q$ CDW charge order is suppressed, $\rho_{2\bm Q} \to 0$  as is the induced charge $4e$ superconducting component, $\Delta_{4e} \to 0$ (from Ref.\cite{Rosales-2024}).}
\label{fig:half-vortex-structure}
\end{figure}

Hence, the PDW has  \textit{three} types of topological textures: 
\begin{enumerate}
\item
Abrikosov vortices with topological charges $(q_s, q_c)=(\pm 1, 0)$ (i.e. with $m_{\bm Q}=m_{-\bm Q}=\pm 1$), and the short-distance behavior  $\lim_{r \to 0} \Delta_{\bm Q}(\bm r)=0$ and $\lim_{r \to 0} \Delta_{-\bm Q}(\bm r)=0$, and carry quantized magnetic flux $\pm 2\pi$.
\item
Double dislocations with topological charges $(q_s, q_c)=(0, \pm 1)$  (i.e. with $m_{\bm Q}=-m_{-\bm Q}=\pm 1$). Here too $\lim_{r \to 0} \Delta_{\bm Q}(\bm r)=0$ and $\lim_{r \to 0} \Delta_{-\bm Q}(\bm r)=0$. Fig. \ref{fig:double} shows a double dislocation.
\item
Half-vortices with topological charges $(q_s, q_c)=(\pm 1/2, \pm 1/2)$ (i.e. with $m_{\bm Q}=\pm 1$ and $m_{-\bm Q}=0$ or $m_{\bm Q}=0$ and $m_{-\bm Q}=\pm 1$), and carry \textit{half-quantized} magnetic flux $\pm \pi$. Hence, half-vortices  are bound to single dislocations with $q_c=\pm 1/2$ Fig. \ref{fig:half} depicts a half-vortex.
For the half vortex with $(q_s, q_c)= (\pm 1/2, \mp 1/2)$, we must have $\lim_{r \to 0} \Delta_{\bm Q}(\bm r)=0$ and $\lim_{r \to 0} \Delta_{-\bm Q}(\bm r)\neq 0$, and similarly for the other cases.  Notice that this implies that the core of the half-vortex of a PDW has FF superconducting order and necessarily breaks inversion symmetry. The structure of the two components of the PDW in the core of a half-vortex is shown in Fig.\ref{fig:half-vortex-structure} (from Ref. \cite{Rosales-2024}).
\end{enumerate}

The existence of half-vortices means that  in the PDW phase the flux quantum is $hc/4e$, which is half of the flux quantum of uniform superconducting states. 
This prediction can be tested experimentally with local SQUID magnetometry. 
The structure of the half-vortex of the unidirectional PDW also has some remarkable features. 
One such feature is that it is bound to a dislocation of unit topological charge and carries $\pi$ flux. 
Another is that the core of the half-vortex has the character of an FF state, which is the state that competes with the PDW state. 
In this sense the half-vortex of a PDW superconductor is a clear manifestation of the concept of intertwined orders. 
In addition, with Marcus Rosales we investigated the effects induced by the different topological textures of the PDW on the electronic states of this superconducting state. 
For details see Ref.\cite{Rosales-2024}.

Topological textures of superconductors with fractional fluxes  have also been discussed in other multicomponent superconductors, such as the kagome superconductors \cite{Ge-2024}, in more general multi-component states \cite{Babaev-2002, Agterberg-2008, Agterberg-2011}, in spinor condensates \cite{Mukerjee-2006}, and in unbalanced Fermi gases \cite{Radzihovsky-2009}. Likewise half-vortices with spin polarization have been proposed in the context of $p_x+ip_y$ superconductors \cite{Vakaryuk-2009}. However we should note that the prediction of superconducting textures carrying fractional magnetic fluxes remains untested experimentally.

 We should also remark that multi-component superconductors with fractional vortices are not necessarily \textit{topological superconductors} which must have Majorana zero modes in their cores. This is expected to happen in the $p_x+ip_y$ superconducting state, which is topological, but not in the PDW which is not a topological superconductor (at least, not necessarily so). In fact, some years ago with Luiz Santos and Yuxuan Wang \cite{Santos-2019} we proposed a $p_x+ip_y$ pair-density wave type state (which is topological) motivated by the experimental observation of nematic orders in close proximity to the paired fractional quantum Hall state in the $N=1$ Landau level.

\begin{figure}[hbt]
\center
\includegraphics[width=0.80\textwidth]{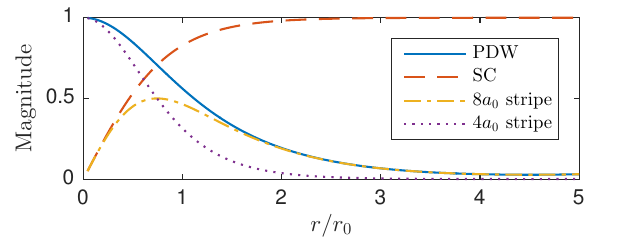}
\caption{Abrikosov vortex halo and PDW order. The dashed red curve is the (normalized) amplitude of the uniform SC order parameter, the uniform blue curve is the (normalized) PDW amplitude, the dashed-dotted orange curve is the amplitude of the period $8a_0$ ($1\bm Q$) of the charge order and the dotted curve is the (normalized) amplitude of the period $4a_0$ ($2\bm Q$) of the charge order. See text for details. (from Ref.\cite{Wang-2018}).}
\label{fig:halo}
\end{figure}  

We now turn to the problem of ``halos'' of superconducting vortices. This problem has a long history in the copper-oxide superconductors \cite{Hoffman-2002,Kivelson-2002}. STM experiments by Edkins and coworkers \cite{Edkins-2018} near a superconducting vortex of {\BSCCO} have made a strong case that the uniform -d-wave superconducting state in {\BSCCO} has a subdominant (competing) PDW state.  Motivated by the STM experiments with Yuxuan Wang and other collaborators \cite{Wang-2018} we considered the problem of a uniform d-wave superconductor which competes with a unidirectional PDW state with wave vector $\bm Q$ (with period $4a_0$). 

Using a Landau-Ginzburg theory with my collaborators \cite{Wang-2018} we investigated the structure of a superconducting vortex. By adjusting a parameter we mapped the solution to finding a meron solution of an $O(3)$ non-linear sigma model. At long distances the solution represents a vortex with unit topological charge while at short distances, where the amplitude of the superconductor order parameter vanishes, a finite PDW component appears (see Fig.\ref{fig:halo}). In this PDW halo, in addition of the period $4a_0$  of the charge order of the PDW (with wave vector $2\bm Q$), there is a period $8a_0$ component of the charge density  (with the same wave vector $\bm Q$ of the PDW) in agreement with the experimental result. In addition the structure of the charge order inside the vortex has tell-tale phase shifts associated with the vorticity in the Fourier transforms of the electronic  tunneling density of states in an idealized STM experiment (see also the related work of Dai et al \cite{Dai-2018}). 
%aqui

%\begin{equation}
%\bm n(\bm r)\sim (\Delta_0(\bm r), 
%\label{eq:NLSM}
%\end{equation}

\subsection{Thermal Phase Transitions and the  Role of Dimensionality}
\label{sec:dimensionality}

We will now discuss the thermal transitions out of the PDW state. The Landau-Ginzburg theory of Eq.\eqref{eq:GL-PDW} is an example of competing orders, in this case between the PDW and an LO phase.
As all such theories, even if improved by state-of-the-art renormalization group approaches \cite{Calabrese-2003}, has the fine-tuning problem emphasized in section \ref{sec:competing}. 

In addition, there is also the problem of understanding the role of composite orders and whether there are phases characterized by composite orders, vestigial orders in the sense of Ref.\cite{Nie-2013} (for a review see Ref.\cite{Fernandes-2019}), exist after the ``mother'' PDW phase has melted. This problem was considered long ago by Golubovi\'c and Kosti\'c \cite{Golubovic-1988} in a suitable large-$N$ limit of  Landau-Ginzburg type  theories with several order parameters. Again, this approach is highly restricted to the close proximity of a rather complex multi-critical point. 

The concept of vestigial order was originally proposed  by Nie, Kivelson and Tarjus \cite{Nie-2013}, in the context of systems with quenched disorder which couple linearly to the ``fundamental'' order parameters thereby destroying such ordered phases leaving behind a phase defined by a composite order parameter (such as the Ising nematic). Mross and Senthil have argued that while disorder may destroy the PDW ling-range order, it may be possible that a uniform charge $4e$ superconductor may be left behind \cite{Mross-2015}, essentially as a form of vestigial order.

Landau-Ginzburg type field theories describe well classical systems near four dimensions (and quantum systems in 3+1 dimensions). However they cannot describe two-dimensional (and quasi-2d) systems which have large fluctuations. It is naturally to explore these questions in a 2d context since the copper oxide superconductors are (mostly) strongly quasi-2d materials. 

\begin{figure}[hbt]
\center
\includegraphics[width=0.50\textwidth]{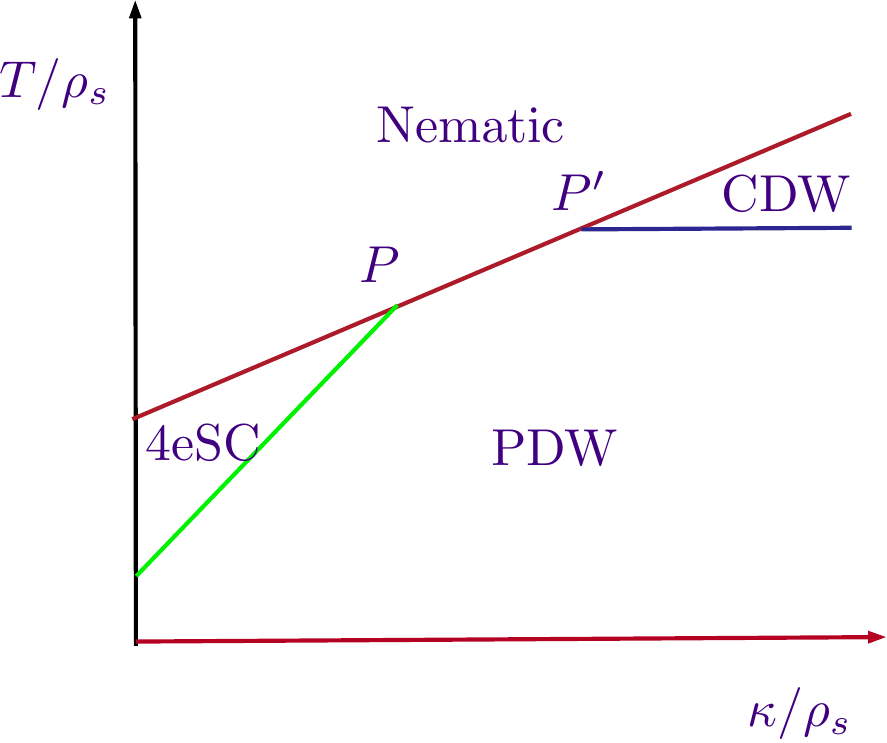}
\caption{Phase diagram of the thermal melting of a unidirectional incommensurate PDW. in two dimensions. Here the red straight line is where the half-vortices proliferate, the black straight line is wheer the vortices proliferate and the orange straight line is where the double dislocations proliferate. All thermal transitions are of the Kosterlitz-Thouless type. The points $P$ and $P'$, where  pairs of lines cross, are multi-critical points with an enhanced $SU(3)_1$ symmetry. See text for details.}
\label{fig:thermal}
\end{figure}

In Ref.\cite{Berg-2009b} with Erez Berg and  Steven Kivelson we considered the (classical) phase diagram of a PDW phase in a  two-dimensional (classical) system with a PDW phase. here I will follow that work closely. Our treatment followed the standard approach of Kosterlitz and Thouless \cite{Kosterlitz-1973} in which the phase transition from the superfluid phase to the normal (symmetric) phase proceeds by thermal proliferation (unbinding) of its topological textures, vortices. Much as in the theory of two-dimensional melting \cite{Halperin-1978,Young-1979}, our phase diagram, shown in Fig.\ref{fig:thermal}, has regimes with partially melted phases (vestigial in the current terminology). However, since the PDW has several topological textures the resulting phase diagram is richer. More significantly, the enhanced fluctuations in 2d to a significant degree circumvent the fine-tuning problem of the Landau-Ginzburg type theories.

The PDW is a state which spontaneously breaks a continuous $U(1) \times U(1)$ global symmetry. In two dimensions this is not allowed by the Mermin-Wagner Theorem. Instead, just as in the case of other systems with a $U(1)$ symmetry, the low temperature phase is not a state of broken symmetry but a critical phase in which, below some critical temperature,  all observables which transform (in this case) under the $U(1) \times U(1)$ symmetry exhibit power law correlation functions. The basic physics is discussed in many textbooks such as in the book of Chaikin and Lubensky \cite{Chaikin-1995} or in my textbooks \cite{Fradkin-FTCMP-2013,Fradkin-QFT-2023}.

The low temperature behavior of low dimensional systems can be described by an effective non-linear sigma model theory in which the magnitude of the order parameter is developed at  relatively high temperatures and the phase transition involves the development of phase coherence at lower temperatures. We will use this formulation and hold the magnitude $\Delta_{\textrm{PDW}}$ of the PDW order parameter and focus on the fluctuations of the phase fields $\theta_{\pm \bm Q}(\bm r)$, or equivalently in terms of the phase fields $\theta(\bm r)$ and $\phi(\bm r)$ defined in Eq.\eqref{eq:PDW-orders}. The effective (classical) Hamiltonian density for the phase fields has the form
\begin{equation}
\mathcal{H}=\frac{1}{2} \rho_s \left(\bm \nabla \theta\right)^2+\frac{1}{2}\kappa  \left(\bm \nabla \phi\right)^2
\label{eq:2d}
\end{equation}
where $\rho_s$ is the superfluid density of the PDW and $\kappa$ is the stiffness of the CDW, and where I have ignored their anisotropy, which does not change the physics in an essential way, but it present in a unidirectional PDW.

Just as in other 2d systems with a global $U(1)$ symmetry, the correlation functions of the order parameters in the PDW phase exhibit power law behaviors
\begin{align}
\Big< \Delta_{\pm \bm Q}(\bm r) \Delta_{\pm \bm Q}^*(\bm r')\Big>=&\frac{\Delta_{\rm PDW}^2}{|\bm r - \bm r'|^{2\Delta(PDW)}}\\
\Big<\rho_{\bm K}(\bm r) \rho_{\bm K}^*\Big>=&\frac{\rho_K^2}{|\bm r - \bm r'|^{2\Delta(CDW)}}\\
\Big< \Delta_{4e}(\bm r) \Delta_{4e}^*(\bm r')\Big>=&\frac{\Delta_{4e}^2}{|\bm r - \bm r'|^{2\Delta(4e)}}
\label{eq:order-parameter-power-laws}
\end{align}
where $\Delta(PDW)$, $\Delta(CDW)$ and $\Delta(4e)$ are the scaling dimensions of the order parameters of the PDW, $\Delta_{\pm \bm Q}$, $\rho_{\bm K}$ and $\Delta_{4e}$, and are respectively  given by
\begin{equation}
\Delta(PDW)=2\pi \left(\frac{T}{\rho_s}+\frac{T}{\kappa}\right), \qquad \Delta(CDW)=\frac{8\pi T}{\kappa}, \qquad \Delta(4e)=\frac{8\pi T}{\rho_s}
\label{eq:scaling-dimensions-op}
\end{equation}
where $T$ is the temperature.

Much as in the Kosterlitz-Thouless theory \cite{Kosterlitz-1973,Jose-1977,Kadanoff-1979} the phase transition out of the PDW phase proceeds by proliferation of its topological excitations. Since there are three types of topological excitations there are different pathways for phase transitions out of the PDW phase. These phase transitions can be described in terms of the \textit{dual fields} $\vartheta$ and $\varphi$ (see, for instance, Ref.\cite{Fradkin-2024}), defined by the Cauchy-Riemann relations
\begin{equation}
\partial_i\theta=\epsilon_{ij}\partial_j \vartheta, \qquad \partial_i\phi=\epsilon_{ij}\partial_j \varphi
\label{eq:CR}
\end{equation}
where $i, j=1,2$ are two orthogonal directions.
The dual Hamiltonian is
\begin{align}
\mathcal{H}_{\rm dual} [\vartheta, \varphi]=&\frac{T}{2\rho_s} (\bm \nabla \vartheta )^2+\frac{T}{2\kappa} (\bm \nabla \varphi )^2\nonumber\\
							      -& g_{1,0} \cos (2\pi \vartheta )-g_{0,1} \cos(2\pi \varphi)-g_{1/2,1/2} \cos(\pi \vartheta) \cos(\pi \varphi)
\label{eq:dual-H}
\end{align}
where the coupling constants $g_{1,0}$, $g_{0,1}$ and $g_{1/2,1/2}$ are, respectively, proportional to the logarithm of the \textit{fugacities} of the vortex operator, 
$\textrm{exp}(\pm i \pi  \vartheta)$, (with topological charge $(1,0)$), the double dislocation operator, $\textrm{exp}(\pm 2\pi i\varphi)$, (with topological charge $(0,1)$) 
and the half-vortex operators, $\textrm{exp} (i\pi \vartheta\pm i \pi \varphi)$ (with topological charge $(1/2/,1/2)$. 
These three topological excitations have an energy cost which scales withe logarithm of the system size. 
The Kosterlitz-Thouless energy-entropy criterion then yields the critical temperatures. However these phase transitions may not necessarily happen at the same temperature.  

The simplest and most direct way to determine qualitatively the phase diagram is by means of a renormalization group analysis of this system, 
which implies that the phase transition happens when an operator becomes marginally relevant. 
An operator is marginal when its scaling dimensions is equal to  the dimensionality of space, two in this case. 
The scaling dimension $\Delta(q_s, q_c)$ of a topological excitation with topological charge $(q_s, q_c)$ is
\begin{equation}
\Delta(q_s, q_c)=\frac{\pi}{T}\left(q_s^2\rho_s+q_c^2 \kappa\right)
\label{eq:topo-dimensions}
\end{equation}
and marginality is achieved if the temperature $T/\rho_s$ and the stiffness $\kappa/\rho_s$ satisfy the condition
\begin{equation}
\frac{\pi}{T}\left(q_s^2\rho_s+q_c^2 \kappa\right)=2
\label{eq:marginality}
\end{equation}

The result of this analysis is presented in Fig.\ref{fig:thermal}. The straight lines where these operators are marginal are generally different. 
In the figure  the half-vortex operator $(\pm 1/2, \mp 1/2)$ becomes marginal at the red line, 
the vortex operator $(\pm 1, 0)$ becomes marginal at the blue line, and the double dislocation operator $(0, \pm 1)$ becomes marginal at the orange line.
Here the incommensurate PDW is the low temperature phase fo all values of $\kappa/\rho_s$.
For values of $\kappa/\rho_s$ to the right of the point denoted by $P'$, with increasing temperature the blue line is reached, 
where the vortices become relevant, the superfluidity is lost. Here the system enters an (again critical) non superfluid phase with  unidirectional incommensurate CDW state 
(with wave vector $2\bm Q$). Eventually, the CDW order is melted at a higher temperature phase transition at the red line. 
This high temperature phase is uniform and anisotropic. This is a Nematic phase which at higher temperatures melts, at an Ising-type phase transition, 
into a uniform and isotropic normal phase. 
On the other hand, for  $\kappa/\rho_s$ between the points $P$ and $P'$ the PDW melts into the high temperature Nematic phase by proliferation of the half-vortices. 
Finally, for $\kappa/\rho_s$ to the left of $P$, the double dislocations proliferate at the orange line and the system enters a uniform phase which is a charge $4e$  
condensate, which in turn  melts into the Nematic phase by proliferating the half-vortices.

The phase diagram of Fig.\ref{fig:thermal} shows that in two dimensions, as a consequence of the existence of large fluctuational regimes,  the fine-tuning problem of competing orders can (and is) be avoided. 

The CDW phase and the charge $4e$ superconducting phase are examples of vestigial phases characterized by the composite operators defined in Eq.\eqref{eq:daughter-CDW} and Eq.\eqref{eq:daughter-4e}, in which some of the (almost) broken symmetries of the PDW are  restored. While the CDW phase is a non-superconducting state with translation symmetry broken in only one direction (with wave vector $2\bm Q$), the charge $4e$ superconducting phase is exotic in the sense that its fundamental vortices are the half-vortices. All these phase transitions are very similar to the conventional Kosterlitz-Thouless (KT) transition. 
Except for the direct PDW-Nematic transition, at which the full $U(1) \times U(1)$ is (almost) spontaneously broken, the other transitions are conventional KT transitions at which only one of the $U(1)$ symmetries is restored. In  Fig.\ref{fig:thermal} , the points $P$ and $P'$ are multi-critical points at which the $U(1) \times U(1)$ symmetry is enhanced (dynamically) and are described by an $SU(3)_1$ conformal field theory \cite{Kondev-1996}.

In the analysis of the PDW phase diagram we ignored a possible uniform superconducting phase. If such a phase exists a term of the form of Eq.\eqref{eq:phase-locking} is allowed in the free energy. This coupling locks the phase fields of the PDW and uniform superconducting orders mod $\pi$. This Ising symmetry is spontaneously broken in the uniform phase. In fact, if $\langle \Delta_0 \rangle \neq 0$, each half-vortex becomes coupled to a  half-anti-vortex by a potential which grows linearly with their separation instead of logarithmically as in the PDW phase. In other words, a half-vortex and an anti-half-vortex are at the endpoints of a domain wall of the spontaneously broken Ising symmetry. The phase transition from the uniform to the PDW phase is an Ising phase transition at which the energy per unit length (the ``string tension'') vanishes. A related problem was considered long ago by Dung-Hai Lee and Geoffrey Grinstein  \cite{DH_Lee-1985} and by Dung-Hai Lee, Geoffrey Grinstein and John Toner \cite{DH_Lee-1986}, and  discussed in depth more recently discussed  by  Yifei Shi, Austin Lamacraft and Paul Fendley \cite{Shi-2011} This competing uniform phase and the uniform-PDW Ising transition is not shown explicitly in Fig. \ref{fig:thermal} where it is assumed to have occurred at much lower temperatures. A hint of this behavior was seen in the lateral Josephson coupling experiments of Ref. \cite{Hamilton-2018}.

\section{Microscopic Models of Intertwined Orders}
\label{sec:microscopic}

As we saw in Section \ref{sec:intertwined}, there is clear experimental evidence for intertwined orders is many materials, not only in the copper oxide superconductors. So far experimental evidence of the PDW, as a prototypical case of intertwined orders, as a new superconducting state of matter is largely restricted to the Lanthanum family of copper oxide superconductors. Nevertheless, as we also saw, evidence for the PDW as a subdominant order has been found in several materials. A striking feature of the PDW is that many experimentally testable predictions follow even at the level of phenomenological theory.

At the level of physically reasonable microscopic models the situation is very different, and much less satisfactory and there is a good reason for that. Microscopic models of intertwined orders are almost without exception analytically quite intractable, and they are challenging even numerically albeit for very different reasons (see, for instance, Ref.\cite{Fradkin-2015}). Here I will briefly cite a few relevant works in the large literature of this still largely open problem.

In spite of my caveats  there are  examples of simple models with intertwined orders, in the sense that we have used this term here, in models of strongly correlated systems that can be studied in a quasi-1D regime where this physics is seen quite clearly (see Refs. \cite{Jaefari-2010,Fradkin-2015}). DMRG studies of models of strongly correlated systems in ladders of various widths have clearly shown a tendency to produce stripe phases and d-wave superconductivity \cite{Noack-1995,White-1998a,Zheng-2017,Arovas-2022}.
The only known examples of models that can be demonstrated to have a PDW phase is the Kondo-Heisenberg chain  \cite{Berg-2010}, and the related two-leg-ladder model of Ref.\cite{Jaefari-2012}, which has a commensurate PDW in much of its phase diagram.

Even the microscopic mean field theory approaches to the PDW have proven challenging. For instance BCS-type mean field  theories of Refs. \cite{Loder-2010} and \cite{Loder-2011} find a PDW only if it has a d-wave form factor and for interactions which are larger than the band-width for the electrons, a strong coupling regime in which the assumptions behind BCS theory do not hold. Other approaches with different microscopic mechanisms have also been proposed \cite{Lee-2014,Wardth-2017}. Quasi-1D approaches also yield PDW phases \cite{Soto-Garrido-2015}.

Variational quantum Monte Carlo studies of the $t-J$ model have suggested that the PDW is a strong competitor of the uniform d-wave superconducting state \cite{Himeda-2002}. A similar result was found with iPEPS tensor networks methods \cite{Corboz-2014,Ponsioen-2023,Verstraete-2008} but with the caveat that they are not consistent with DMRG computations in relatively wide cylinders. There is evidence for PDW phases in one-dimensional systems with strong electron-phonon coupling \cite{Huang-2022} and in the (three-band) Emery model \cite{Jiang-2023}. Finally, we should note that nematic phases in the spin-triplet channel \cite{Wu-2007} (``altermagnets'') naturally yield complex phase diagrams with PDW superconducting states \cite{Soto-Garrido-2014} and p-wave triplet phases with  time reversal symmetry breaking \cite{zou-2025}.

Phil Anderson's proposal of a novel possible ground state of quantum antiferromagnets, which he dubbed a \textit{resonating valence bond} state (RVB) \cite{Anderson-1987} was the first of these new approaches and had a profound impact in the field. It articulated the fact that the new superconducting states resulted from doping a \textit{Mott insulator}, and the related later work by  Kivelson, Rokhsar and Sethna \cite{Kivelson-1987} which assumed that the valence bonds may have support on just bonds between nearest neighbors, which can be represented as ground states  of quantum dimer models \cite{Rokhsar-1988}.  A feature of these proposals was the notion that the ground state of a quantum antiferromagnet may not be a state with a spontaneously broken global symmetry of a spin system but it  instead may be a \textit{spin liquid}, a translationally invariant state of spin-1/2 quantum antiferromagnets whose low energy excitations carry the spin 1/2 quantum number. 
It was later realized that spin liquids are  (or are closely related to) a topological state of matter which became  a central problem of Condensed Matter Physics.   

An aspect of these  approaches was that they naturally required the introduction of ideas of lattice gauge theories \cite{Baskaran-1988} which inspired my subsequent work with Kivelson in which we reformulated quantum dimer models in the language of gauge theory  \cite{Fradkin-1988b,Fradkin-1990}. Nevertheless it soon became clear that  {\LCO} is naturally well described by a spin-1/2 quantum antiferromagnet on square (and hence \textit{bipartite}) lattice whose ground state is N\'eel state which is not a spin liquid and breaks spontaneously its global SU(2) symmetry. Later work on quantum dimer models by Moessner and Sondhi \cite{Moessner-2001,Moessner-2001c} revealed that \textit{quantum frustration} in non-bipartite lattices was essential for the existence of such a spin liquid state, a topological phase known as  the  $\mathbb{Z}_2$ spin liquid.

The fact that we have to deal with strongly interacting systems led to the exploration of new approaches such as the large-$N$ and large-$S$ treatments of quantum antiferromagnets by Affleck and Marston \cite{Affleck-1988} and by Read and Sachdev \cite{Read-1989c,Read-1991,Sachdev-1991}. These novel approaches, which required the generalization of the global symmetry from $SU(2)$ to $SU(N)$ and the consideration of representations higher than the fundamental, led to the discovery of flux phases and valence bond crystals. Wen, Wilczek and Zee showed that such models naturally lead  to a possible chiral spin liquid which spontaneously breaks time reversal invariance \cite{Wen-1989}, and that had been conjectured by Kalmeyer and Laughlin \cite{Kalmeyer-1987} using an analogy withe the Laughlin wave function for a fractional quantum Hall state of \textit{bosons} (see the recent numerical work by Huang, Gong and Sheng \cite{Huang-2023}). The large-$N$ formulation was also suitable to include the effects of hole-doping and to a mean field theories of d-wave superconductivity, as in the work by Lee, Nagaosa and Wen \cite{Lee-2006}, although requiring the use of somewhat uncontrolled approximations. Recent numerical DMRG studies have confirmed that lightly doped spin liquid ground states are indeed superconductors \cite{Jiang-2021}.

%%%%%%%%%%%%%%%%%%%%%%%%%%%%%%%%%%%%%%%%%%
\section{Discussion and Conclusions}

This paper is an expanded version of the content of the talk that I gave on occasion of being awarded the 2024 Eugene Feenberg Memorial Medal at the
 \textit{Recent Progress in Many-Body Theories} conference  (RPMBT22) held at the Kasuga Campus of the University of Tsukuba (Tsukuba, Japan) on  23-27 September, 2024. It mainly covers the work that I have done in the past two decades on the subject on Intertwined Orders with many collaborators, and I have not discussed my many other contributions in other areas of Condensed Matter Physics.  Although I have included many references to work done by other people, it is not an exhaustive review. I apologize if I have not covered fairly enough work by many other authors in this important area of research in Condensed Matter Physics.

\acknowledgments{My work on this problem has benefitted enormously from the collaboration and insights of many people, both experimentalists and theorists. I am particularly grateful and indebted to Steven Kivelson with who I had the privilege to have shared the  development of these ideas which constitute a reformulation of this problem. I have also benefitted from the collaboration with several theorists including Erez Berg, Rodrigo Soto-Garrido and Yuxuan Wang, as well as many experimentalists, particularly John Tranquada, Dale Van Harlingen (who passed away in 2024), Peter Abbamonte, J. C. S\'eamus Davis, Vidya Madhavan and many others. My work is supported in part by the National Science Foundation grant No. DMR 2225920 at the University of Illinois. }

%%%%%%%%%%%%%%%%%%%%
%\bibliography{feenberg.bib}

%%%%%%%%%%%%%%%%%%%%%%
\end{document}